\documentclass[twocolumn]{aastex631}

\usepackage{mathtools}
\usepackage{amssymb}
\usepackage{appendix}
\usepackage{comment}
\shorttitle{AASTeX v6.3.1 Sample article}
\shortauthors{Qiao et al.}
\graphicspath{{./}{figures/}}

\begin{document}
\title{A Tale of Two Origins: In-Situ versus Accreted Nitrogen-Rich Field Stars in the MW\\}

\author{Yi Qiao}
\author{Baitian Tang}
\affiliation{School of Physics and Astronomy, Sun Yat-sen University, Daxue Road, Zhuhai, 519082, P.R.China; tangbt@mail.sysu.edu.cn}
\affiliation{CSST Science Center for the Guangdong-Hong Kong-Macau Greater Bay Area, Zhuhai, 519082, P.R.China}

\author{Jos\'{e} G. Fern\'{a}ndez-Trincado}
\affiliation{Universidad Cat\'olica del Norte, N\'ucleo UCN en Arqueolog\'ia Gal\'actica - Inst. de Astronom\'ia, Av. Angamos 0610, Antofagasta, Chile}
\affiliation{Universidad Cat\'olica del Norte, Departamento de Ingenier\'ia de Sistemas y Computaci\'on, Av. Angamos 0610, Antofagasta, Chile}

\author{Mingjie Jian}
\affiliation{Department of Astronomy, AlbaNova University Center, Stockholm University, SE–106 91 Stockholm, Sweden}

\author{Carlos Allende Prieto}
\affiliation{Instituto de Astrofísica de Canarias, C/Via Lactea s/n, 38205 La Laguna, Tenerife, Spain}
\affiliation{Departamento de Astrof\'{i}sica, Universidad de La Laguna, 38206 La Laguna, Tenerife, Spain}

\author{Hongliang Yan}
\affiliation{CAS Key Laboratory of Optical Astronomy, National Astronomical Observatories, Chinese Academy of Sciences, Beijing 100012, P.R.China}

\author{Zhen Yuan}
\affiliation{School of Astronomy and Space Science, Nanjing University, Key Laboratory of Modern Astronomy and Astrophysics (Nanjing University), Ministry of Education, Nanjing 210093, P.R.China}
\affiliation{Key Laboratory of Modern Astronomy and Astrophysics, Nanjing University, Ministry of Education, Nanjing 210093, China}

\author{Yang Huang}
\affiliation{CAS Key Laboratory of Optical Astronomy, National Astronomical Observatories, Chinese Academy of Sciences, Beijing 100012, P.R.China}

\author{Thomas Masseron}
\affiliation{Instituto de Astrofísica de Canarias, C/Via Lactea s/n, 38205 La Laguna, Tenerife, Spain}
\affiliation{Departamento de Astrof\'{i}sica, Universidad de La Laguna, 38206 La Laguna, Tenerife, Spain}

\author{Beatriz Barbuy}
\affiliation{Universidade de São Paulo, IAG, Departamento de Astronomia, 05508-090 São Paulo, Brazil}

\author{Jianrong Shi}
\affiliation{CAS Key Laboratory of Optical Astronomy, National Astronomical Observatories, Chinese Academy of Sciences, Beijing 100012, P.R.China}

\author{Chengyuan Li}
\author{Ruoyun Huang}
\author{Jiajun Zhang}
\affiliation{School of Physics and Astronomy, Sun Yat-sen University, Daxue Road, Zhuhai, 519082, P.R.China}
\affiliation{CSST Science Center for the Guangdong-Hong Kong-Macau Greater Bay Area, Zhuhai, 519082, P.R.China}

\author{Jing Li}
\affiliation{School of Physics and Astronomy, China West Normal University, 1 ShiDa Road, Nanchong, 637002, P.R.China}

\author{Chao Liu}
\affiliation{Key Laboratory of Space Astronomy and Technology, National Astronomical Observatories, Chinese Academy of Sciences, Beijing 100101, P.R.China}
\affiliation{Institute for Frontiers in Astronomy and Astrophysics, Beijing Normal University, Beijing 102206, P.R.China}

\author{Weishan Zhu}
\affiliation{School of Physics and Astronomy, Sun Yat-sen University, Daxue Road, Zhuhai, 519082, P.R.China}
\affiliation{CSST Science Center for the Guangdong-Hong Kong-Macau Greater Bay Area, Zhuhai, 519082, P.R.China}

\received{XXX} 
\submitjournal{The Astrophysical Journal}

\begin{abstract}

$\hspace*{0.3cm}$Spectroscopic surveys have identified significant numbers of metal-poor nitrogen-rich (N-rich) field stars. These stars are strong candidates for escapees from globular clusters (GCs), as their distinctive nitrogen enhancement mirrors the chemical patterns observed in some of the members of GCs. As part of the effort to characterize their chemodynamical properties, we derived abundances for up to 25 elements in a sample of 33 N-rich field giant stars (18 of them are studied for the first time) using high-resolution optical spectroscopy. We confirm their elevated abundances of N, Na, and Al, strongly supporting a GC origin. Given that Galactic GCs themselves formed within diverse progenitor galaxies, we sought to identify the ancestral systems of these N-rich field stars. By analyzing their dynamical parameters, we separated the sample into high-energy (HE) and low-energy (LE) groups. The HE group exhibits lower [$\alpha$/Fe] and enhanced {\it r}-process abundances compared to the LE group. This indicates that the HE stars likely escaped from GCs accreted from massive dwarf galaxies (e.g., Gaia-Sausage-Enceladus), while the LE stars probably originated from in-situ GCs. We also find that the chemical pattern of these N-rich stars with [Fe/H$] \lessapprox -1.0$ are similar to the high-redshift ``N-emitters''. Furthermore, orbital integrations revealed a close encounter between one N-rich field star and the globular cluster NGC 6235. Our work demonstrates the potential of using chemodynamical analyses to trace Galactic assembly through chemical peculiar stars, while highlighting that larger samples and more precise data in the future are crucial to establish definitive origins. 

\end{abstract}


\section{Introduction}
\label{sec:1}

$\hspace*{0.2cm}$The Milky Way (MW) represents a archetypal laboratory for studying galaxy formation through hierarchical assembly. Seminal studies established that the MW halo preserves fossil records of accretion events, where tidally disrupted dwarf galaxies (Sagittarius) imprint kinematical and chemical substructures \citep{sgr_discov_1994Natur.370..194I,sgr1_2003,sgr_chemic_2007A&A...465..815S,2021ApJ...923..172H}. This paradigm was revolutionized by Gaia astrometry and panoramic spectroscopic surveys (e.g., Sloan Digital Sky Survey --- SDSS, \citealt{Blanton2017}; LAMOST Galactic spectroscopic survey, \citealt{cui_2012RAA....12.1197C,Deng2012,Zhao2012}). The newly identified Gaia-Sausage-Enceladus (GSE) merger is an accretion event that contributed a substantial amount of mass to the inner halo \citep{gse_belo_2018MNRAS.478..611B,helmi_Natur.563...85H,gse1_2018,gse2_2019}, while the MW's stellar disk shows dynamical signatures of disequilibrium, such as phase spirals \citep{Antoja2018}, spiral arms, and ripples \citep{Laporte2020,Qiao2024}. Earlier studies \citep{halo1_2010,NS10,dual_halo_2012ApJ...746...34B} had already identified the halo’s dual structures, inner (high-[$\alpha$/Fe]) and outer (low-[$\alpha$/Fe]) components, reflecting distinct accretion histories. Globular clusters (GCs) serve as critical tracers since they share the dynamical and chemical signatures of their host accretion systems \citep {Massari2019,LinSH2025}, bridging galactic and cluster-scale evolution. 

\begin{table*}[htbp]
	\raggedright
	\caption{Stellar Parameters and Chemical Abundances for 8 N-rich Field Stars Observed with CFHT. }
	\label{tab:1}
	\begin{tabular}{c c c c c c c c c}
		\hline\hline\noalign{\smallskip}	
		Object & Num28 & Num32 & Num24 & Num7 & Num63 & Num1 & Num37 & Num10 \\
		\noalign{\smallskip}\hline\noalign{\smallskip}
		RA (J2000) & 13.1551 & 140.2051 & 115.7504 & 145.2212 & 239.9169 & 260.9612 & 180.2667 & 286.7982 \\
		Dec (J2000) & 37.6988 & 32.4165 & 15.5836 & 25.7415 & 32.446 & 49.5797 & $-$0.9752 & 39.1366 \\
		$\mathrm{T_{eff}}$ (K) & 4775 & 5065 & 5271 & 4283 & 4558 & 4707 & 4920 & 4458 \\
		log $\mathrm{g}$ & 1.57 & 2.56 & 1.95 & 1.05 & 1.16 & 1.70 & 2.25 & 1.33 \\
		$\mathrm{[Fe/H]}$ & $-$1.060 & $-$1.245 & $-$1.260 & $-$1.125 & $-$1.790 & $-$1.370 & $-$1.390 & $-$1.110 \\
		$\mathrm{V_{mic}}$ (km $\mathrm{s^{-1}}$) & 1.74 & 1.32 & 1.71 & 2.00 & 1.80 & 1.75 & 1.57 & 2.12 \\
		RV (km $\mathrm{s^{-1}}$) & 3.93 & $-$1.44 & $-$139.73 & 207.00 & $-$141.95 & $-$207.05 & 134.62 & $-$312.34 \\
		$\mathrm{[O/Fe]}$ & 0.57 & 0.58 & 0.48 & 0.54 & 0.06 & 0.31 & 0.52 & 0.56 \\
		$\mathrm{[Mg/Fe]}$ & 0.64 & 0.37 & 0.55 & 0.30 & 0.47 & 0.59 & 0.50 & 0.55 \\
		$\mathrm{[Si/Fe]}$ & 0.34 & 0.28 & 0.50 & 0.20 & 0.39 & 0.33 & 0.38 & 0.23 \\
		$\mathrm{[Ca/Fe]}$ & 0.37 & 0.31 & 0.30 & 0.19 & 0.45 & 0.40 & 0.46 & 0.36 \\
		$\mathrm{[Ti/Fe]}$ & 0.44 & 0.24 & 0.27 & 0.36 & 0.28 & 0.35 & 0.33 & 0.50 \\
		$\mathrm{[Na/Fe]}$ & 0.72 & 0.11 & -0.06 & -0.02 & 0.50 & 0.02 & -0.041 & 0.36 \\
		$\mathrm{[Al/Fe]}$ & 0.53 & 0.28 & 0.64 & 0.26 & 1.12 & 0.33 & 0.32 & 0.28 \\
		$\mathrm{[Sc/Fe]}$ & 0.28 & 0.06 & 0.14 & 0.25 & 0.03 & 0.21 & 0.28 & 0.08 \\
		$\mathrm{[Cr/Fe]}$ & 0.04 & $-$0.09 & $-$0.07 & $-$0.02 & 0.02 & 0.04 & 0.00 & 0.12 \\
		$\mathrm{[Mn/Fe]}$ & 0.10 & $-$0.27 & $-$0.34 & $-$0.28 & $-$0.27 & $-$0.14 & $-$0.32 & $-$0.13 \\
		$\mathrm{[Co/Fe]}$ & 0.49 & 0.01 & $-$0.23 & $-$0.04 & $-$0.03 & $-$0.08 & $-$0.13 & $-$0.01 \\
		$\mathrm{[Ni/Fe]}$ & 0.15 & $-$0.02 & 0.06 & $-$0.06 & $-$0.04 & 0.08 & 0.07 & 0.00 \\
		$\mathrm{[V/Fe]}$ & 0.27 & 0.03 & 0.05 & 0.22 & 0.11 & 0.24 & 0.15 & 0.39 \\
		$\mathrm{[Ba/Fe]}$ & 0.48 & 0.74 & 1.02 & 0.47 & $-$0.14 & $-$0.26 & 0.97 & 0.78 \\
		$\mathrm{[La/Fe]}$ & 0.23 & 0.35 & 0.47 & 0.76 & 0.26 & 0.34 & 1.02 & 1.06 \\
		$\mathrm{[Ce/Fe]}$ & 0.01 & 0.21 & 0.22 & 0.30 & 0.00 & 0.06 & 0.76 & 0.52 \\
		$\mathrm{[Eu/Fe]}$ & 0.34 & 0.45 & 0.46 & 0.94 & -- & 0.71 & 0.62 & 1.14 \\
		$\mathrm{[Y/Fe]}$ & $-$0.18 & 0.14 & 0.16 & $-$0.11 & $-$0.30 & $-$0.22 & 0.59 & 0.05  \\
		$\mathrm{[Nd/Fe]}$ & 0.16 & 0.23 & 0.23 & 0.71 & 0.21 & 0.28 & 0.85 & 0.78 \\
		$\mathrm{[Zr/Fe]}$ & 0.29 & 0.29 & 0.70 & 0.54 & 0.23 & 0.23 & 0.94 & -- \\
		\noalign{\smallskip}\hline
	\end{tabular}
\end{table*}

$\hspace*{0.2cm}$The discovery of multiple populations (MPs) in GCs significantly altered our understanding of star formation environments. \cite{2009A&A...505..139C,2009A&A...505..117C} demonstrated that light elements' variations (e.g., Na-O anticorrelations) are ubiquitous in GCs. It is suggested such variations arise from high temperature (T $>7\times{10^7}$ K) proton-capture nucleosynthesis, possibly occurring in early generations of massive stars, such as fast-rotating massive stars (FRMS), supermassive stars, or AGB stars \citep{mps_1_2011,mps_2_2011A&A...533A..69C,mps_3_2018ARA&A..56...83B}. MPs may originate from self-enrichment by these stars, polluting the intracluster medium with processed material. A more metal-poor, and compact environment is preferred for the formation of MPs in GCs \citep{Krause2016,Huang2024}. Dynamical modeling indicates that tidal interactions with the MW stripped more than 50 percent of GC stars into the halo over cosmic time \citep{2003_simulation_gc}, and escaped stars should retain MPs signatures, including N, Na, and Al enrichment, enabling their identification as relics of dissolved clusters \citep{lind_2015A&A...575L..12L,martell_2016ApJ...825..146M,nrich_1_2017MNRAS.465..501S}. 

$\hspace*{0.2cm}$Among the aforementioned enriched elements in GCs, N is easier to estimate through molecular bands (e.g., CN bands around 3883 and 4215 \AA). Therefore, N-rich field stars have emerged as powerful tracers of GC tidal disruption and accretion events within the MW. Low-resolution surveys (SDSS/SEGUE, LAMOST) firstly identified such stars through CN-band enhancements, associating them with escaped GC second generations \citep{nrich_marttell_2011A&A...534A.136M, tang2019, yan_lamost_2022Innov...300224Y}. APOGEE spectroscopy later confirmed their distinct chemistry: anomalous [C/N], [Al/Fe], and [Mg/Fe] ratios mirroring features of GC enhanced populations, indicative of proton-capture processes in dense cluster environments \citep{nrich_1_2017MNRAS.465..501S, FT2017, FT2019, jose_2022A&A...663A.126F}. Detailed studies further revealed kinematic subgroups: while most align with in-situ halo dynamics, 15–30 percent exhibit high-eccentricity ``sausage-like'' orbits characteristic of GES-like mergers \citep{tang2020,kisku_2021MNRAS.504.1657K,yu2021}. Kinematic data and dynamical modeling also enabled their associations with specific GCs \citep{Savino2019, Xu2024}.

$\hspace*{0.2cm}$Despite progress mentioned above, more comprehensive and detailed chemodynamical patterns of N-rich stars, as well as more robust connections with GCs, are still lacking. Similar dynamics (e.g., overlaps in Toomre diagrams) alone cannot distinguish escape events from co-natal stars in accreted systems. Meanwhile, limited observational constraints of heavier elements hinder origin discrimination and r/s-process diagnostics of host environments. These gaps impede a unified understanding of how cluster dissolution and hierarchical assembly collectively shape the N-rich field star populations. 

$\hspace*{0.2cm}$An important preceding work, \cite{yu2021}, confirmed that the chemical abundances (e.g., Na, Al, $\alpha$ elements) of 15 N-rich field stars are similar to GC stars using high-resolution spectroscopy and identified one star with chemical signatures indicative of a GSE origin. Our work significantly expands the previous one by analyzing a larger sample of 33 N-rich field stars (including the 15 stars mentioned above) with high-resolution optical spectra (Section \ref{sec:2}) to study more species and conducting a systematic orbital analysis. We classify the N-rich stars into High-Energy (likely accreted) and Low-Energy (likely in-situ) groups, revealing clearer chemical distinctions between them (such as r/s-process ratios), thereby more precisely tracing their origins (Section \ref{sec:3}). Furthermore, in Section \ref{sec:4}, we link metal-poor N-rich field stars to high-redshift ``N-emitters'' and, through orbital integration, provide evidence for an N-rich star possibly ejected from the globular cluster NGC 6235, marking methodological and scientific progress. A brief summary is given in Section \ref{sec:5}. 

\section{Data and Methods}
\label{sec:2}

$\hspace*{0.2cm}$High-resolution spectra were obtained for 23 N-rich field stars selected from \citet{tang2019, tang2020}, in which the original selection from LAMOST was based on a direct comparison of spectral indices measuring the strength of molecular bands. N-rich candidates were identified by anomalously strong CN bands (at $\sim$ 3883 \AA$ $ and $\sim$ 4215 \AA) relative to a normal CH band (at $\sim$ 4300 \AA) \citep{cnbands_2003AJ....125..197H}. This differential index approach effectively isolated stars with enhanced nitrogen without requiring absolute abundance measurements from the low-resolution spectra. About the 23 N-rich field stars: (1) 8 of them were observed with the ESPaDOnS (Echelle SpectroPolarimetric Device for the Observation of Stars, \citealt{2006ASPC..358..362D}) on the Canada-France-Hawaii Telescope (CFHT). This spectrograph offers high-resolution spectra with large wavelength coverage ($R=68000$, $3700<\lambda<10500$ $\mathrm{\AA}$). These spectra were obtained between November 2020 and July 2023 as part of the Telescope Access Program. The signal-to-noise ratios (SNR) per pixel typically exceed 60 between 4300 - 8600 \AA. (2) 15 of them were observed with the MIKE (Magellan Inamori Kyocera Echelle, \citealt{2003SPIE.4841.1694B}) on the Magellan Clay telescope. The MIKE spectra achieve nominal spectral resolution of 35000/28000 on the blue side ($3200<\lambda<5000$ \AA) and the red side ($4900<\lambda<10000$ \AA), respectively. \citet{yu2021} have presented an analysis of 21 elemental abundances of these 15 stars. In addition, we derive the abundances of C, N, Cu, and Zn in this work. 

\begin{table}[htbp]
	\centering
	\caption{Chemical Abundances of C, N, Cu and Zn for 23 LAMOST-selected N-rich Field Stars.}
	\label{tab:2}
	\begin{tabular}{c c c c c c c}
		\hline\hline\noalign{\smallskip}	
		Object & [C/Fe] & [N/Fe] & [Cu/Fe] & [Zn/Fe] \\
		\noalign{\smallskip}\hline\noalign{\smallskip}
		$\mathrm{Num28}$ & 0.15 & 1.09 & 0.02 & 0.37 \\
		$\mathrm{Num32}$ & 0.12	& 1.00 & $-$0.33 & 0.24 \\
		$\mathrm{Num24}$ & 0.05	& 0.78 & $-$0.46 & 0.13 \\
		$\mathrm{Num7}$ & $-$0.41 & 0.78 & $-$0.95	& 0.00 \\
		$\mathrm{Num63}$ & $-$0.59 & 1.06 & $-$0.65 & 0.14 \\
		$\mathrm{Num1}$ & $-$0.34 & 0.84 & $-$0.70 & 0.03 \\
		$\mathrm{Num37}$ & $-$0.15 & 1.31 & $-$0.48 & 0.03 \\
		$\mathrm{Num10}$ & $-$0.19 & 0.73 & $-$0.40	& 0.27 \\
		$\mathrm{Num9}$ & -- & -- & $-$0.85 & $-$0.38 \\
		$\mathrm{Num38}$ & $-$0.61 & 1.59 & $-$0.47 & 0.27 \\
		$\mathrm{Num47}$ & $-$0.24 & 1.41 & $-$0.42 & 0.00 \\
		$\mathrm{Num58}$ & $-$0.21 & 1.66 & $-$0.49 & 0.07 \\
		$\mathrm{Num62}$ & $-$0.59 & 1.72 & $-$0.79 & 0.04 \\
		$\mathrm{Num65}$ & -- & -- & $-$0.27 & $-$0.09 \\
		$\mathrm{Num67}$ &  -- & -- & $-$0.19 & 0.25 \\
		$\mathrm{Num69}$ & -- & -- & 0.25 & 0.22 \\
		$\mathrm{Num80}$ & $-$0.12 & 1.52 & $-$0.59 & 0.22 \\
		$\mathrm{Num82}$ & $-$0.14 & 0.84 & $-$0.33 & 0.27 \\
		$\mathrm{Num86}$ & -- & -- & $-$0.42 & $-$0.22 \\
		$\mathrm{Num88}$ & -- & -- & $-$0.55 & 0.15 \\
		$\mathrm{Num94}$ & 0.11 & 1.21 & $-$0.22 & 0.17 \\
		$\mathrm{Num97}$ & -- & -- & $-$0.27 & 0.34 \\
		$\mathrm{Num98}$ & -- & -- & $-$0.01 & 0.16 \\
		\noalign{\smallskip}\hline
	\end{tabular}
\end{table}
 
To further enlarge the N-rich field star sample with optical high-resolution spectra, we crossmatched the APOGEE-selected N-rich field star sample \citep{jose_2022A&A...663A.126F} with the GALAH Data Release (DR) 4 \citep{galah_dr4}. Note that the metallicity cut in the APOGEE N-rich field stars is [$-$1.8, $-$0.7], which is different than that of LAMOST-selected stars ([$-$1.8, $-$1.0]). The GALAH DR4 spectra, obtained using the HERMES spectrograph, cover four distinct wavelength bands (blue: 4675–4949 \AA; green: 5624–5900 \AA; red: 6424–6775 \AA; and near-infrared: 7549–7925 \AA) with a resolution of R $\sim$ 28,000. 10 stars with high SNR (SNR $> 50$ for blue and green bands, SNR $> 75$ for red and NIR bands) were selected. In summary, we collected 33 ($8+15+10$) N-rich field stars (their observation information can be found in Table \ref{tab:a2}) with optical high-resolution spectra for subsequent analysis. 

$\hspace*{0.2cm}$All ESPaDOnS spectra were reduced using standard pipelines, including bias correction, flat-fielding, cosmic ray removal, and wavelength calibration. Next, heliocentric radial velocities (RVs) were derived and corrected using iSpec \citep{2014A&A...569A.111B}. Our derived RVs are consistent with those published by Gaia DR3 within 2 km s$^{-1}$. We didn't find evidence of binaries after comparing RVs of different epochs ($\Delta<2$ km s$^{-1}$). 

$\hspace*{0.2cm}$The stellar atmospheric parameters and chemical abundances were derived using BACCHUS (Brussels Automated Code for Characterizing High accUracy Spectra) \citep{bacchus_2016}, which is developed on top of the radiative transfer code Turbospectrum \citep{2012ascl.soft05004P}. This procedure mainly follows \cite{yu2021} for atomic chemical abundances and \cite{Tang2023} for chemical abundances derived from molecular lines (e.g., C \& N). We briefly recap the main steps here. The model atmospheres were interpolated from MARCS model grids \citep{Gustafsson2008}. The stellar parameters --- effective temperature ($\mathrm{T_{eff}}$), surface gravity (log g), metallicity ([Fe/H]) and microturbulence --- were determined iteratively by enforcing excitation and ionization equilibrium for Fe I and Fe II lines. BACCHUS performs a line-by-line analysis, where observed spectra are compared with synthetic spectra of various abundances. The final abundance of a given element was derived by averaging the best-fit abundances from reliable multiple lines (BACCHUS gives abundance flags of each line and in this work we only chose the abundance values with flag = 0 or checked the lines by eye). 

$\hspace*{0.2cm}$We determined the abundances of C \& N in a self-consistent way. Besides atomic lines, multiple molecular lines are used here. C abundance is firstly derived from the CH lines, then N abundance is derived from the CN lines\footnote{These are not the same feature lines that we used to select N-rich field stars from LAMOST.} (see Fig.\ref{figspec} for sample spectra). The procedure is repeated until the derived abundances between two adjacent iterations converge ($<0.1$ dex). This process can minimize the dependence of the molecular lines. However, if the derived abundances do not converge after 50 iterations, they are not reported in this work. The non-convergence of C/N abundances for several stars is probably attributed to various reasons, such as locally lower S/N at specific wavelengths, limitations of the 1D atmospheric model grid, or intrinsic complexities in the spectra, rather than the absence of N-enhancement. Despite the non-convergence, these stars still exhibit strong CN bands and lines in their spectra. For the 10 stars in our GALAH/APOGEE sample, given that most C/N related spectral lines are absent or noisy in their optical GALAH spectra, we adopted the [C/Fe] and [N/Fe] derived by the APOGEE Stellar Parameter and Chemical Abundance Pipeline \citep[ASPCAP,][]{2016AJ....151..144G} after checking their quality flags (ASPCAPFLAG = 0, C\_FE\_FLAG = 0, N\_FE\_FLAG = 0). 

$\hspace*{0.2cm}$In summary, we determined abundances for 24 species, including light elements (C, N, O), odd-Z elements (Na, Al), $\alpha$-elements (Mg, Si, Ca, Ti), iron-peak elements (Sc, V, Cr, Mn, Co, Ni), neutron-capture elements (Y, Zr, Ba, La, Ce, Nd, Eu) and Cu, Zn. 
This approach ensures high-precision results suitable for studying the chemical evolution of N-rich field stars.  The derived chemical abundances are given in Table \ref{tab:1}, \ref{tab:2} \& \ref{tab:a1}. 

\section{Results}
\label{sec:3}

\subsection{Chemical abundances}
\label{sec:abu}

\begin{figure}[htbp]
    \raggedright
    \includegraphics[width=8.5cm,height=7.9cm]{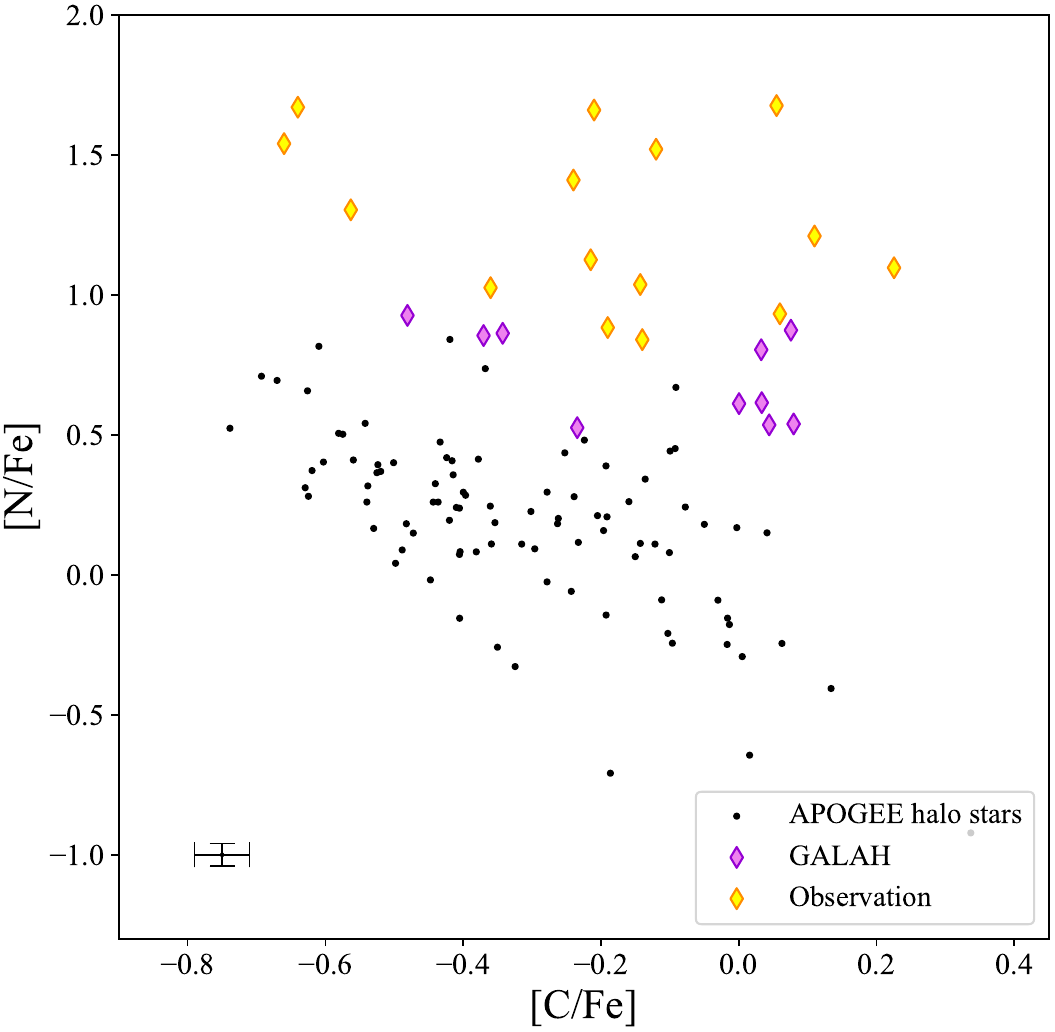}
	\caption{The [N/Fe] versus [C/Fe] plane. Yellow diamonds denote our observed N-rich field stars, while violet diamonds represent 10 GALAH/APOGEE N-rich field stars. The black dots indicate metal-poor RGB halo stars with similar stellar parameters as these N-rich field stars, and they are selected from the APOGEE survey. Typical abundance errors are shown in the corners of abundance panels. }
	\label{figcn}
\end{figure}

\begin{figure}[htbp]
	\raggedright
	\includegraphics[width=8.38cm,height=6.4cm]{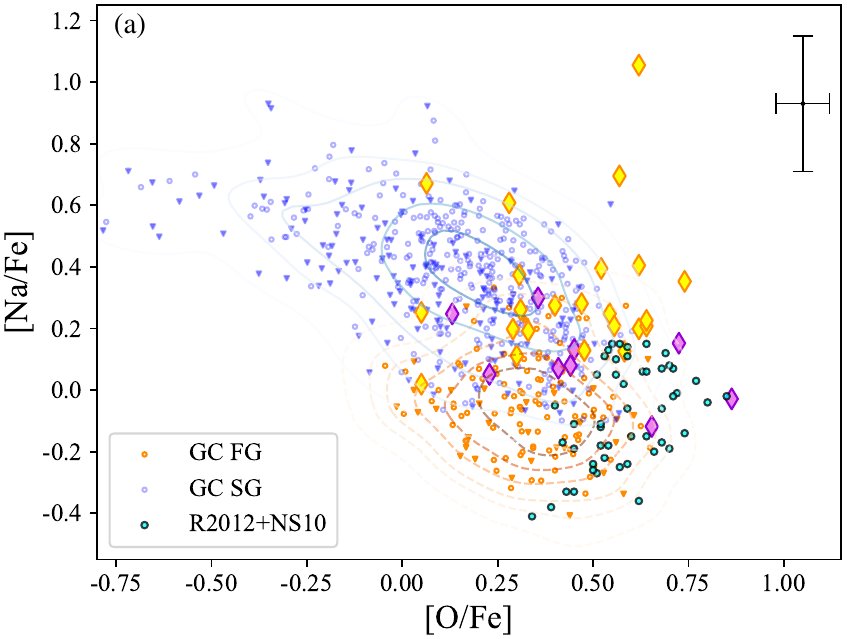}
	\includegraphics[width=8.45cm,height=6.4cm]{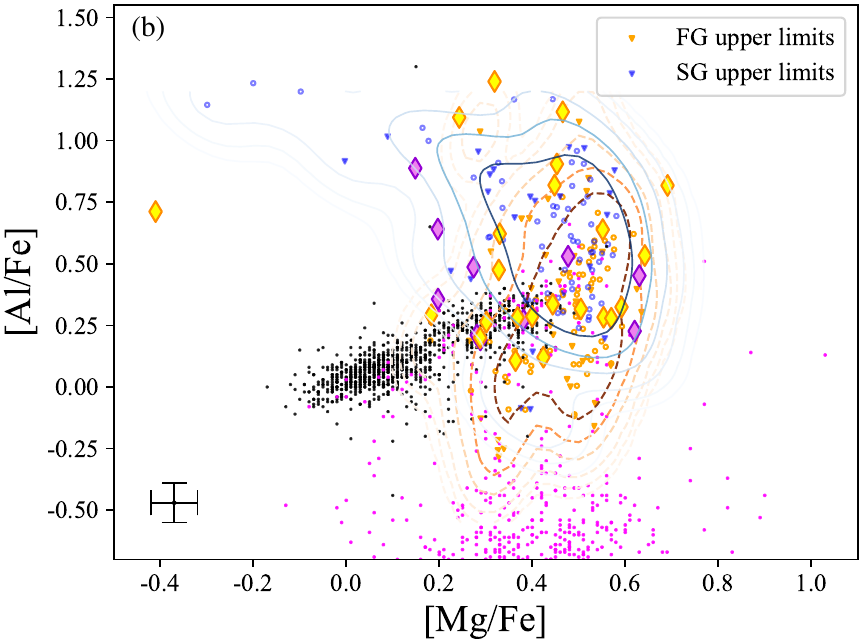}
	\caption{Panel a: The [Na/Fe] - [O/Fe] plane. Panel b: The [Mg/Fe] - [Al/Fe] plane. Markers of N-rich field stars are same as Fig.\ref{figcn}. The Na and O abundances of individual stars in the GCs from \cite{2009A&A...505..117C} are over-plotted for comparison. Orange and blue symbols correspond to GC first generation (FG) and second generation (SG) stars respectively. Upper limits in O and Al abundances are shown as triangles, while detections are shown as dots. The solid and dashed contours indicate the number densities of FG and SG, respectively. The field stars from \cite{ramirez-2012} and \cite{NS10} shown as small cyan circles. The magenta dots correspond to MW halo stars \citep{halo-fulbright-2000AJ....120.1841F, halo-cayrel-2004A&A...416.1117C, halo-barklem-2005A&A...439..129B, halo-yong-2013ApJ...762...26Y, halo-Roederer-2014AJ....147..136R} and black dots are disk stars \citep{disk-reddy-2003MNRAS.340..304R, disk-reddy-2006MNRAS.367.1329R, disk-bensby-2014A&A...562A..71B}. }
	\label{fignao}
\end{figure}

\begin{figure*}[htbp]
	\centering
	\includegraphics[width=17.2cm,height=13.2cm]{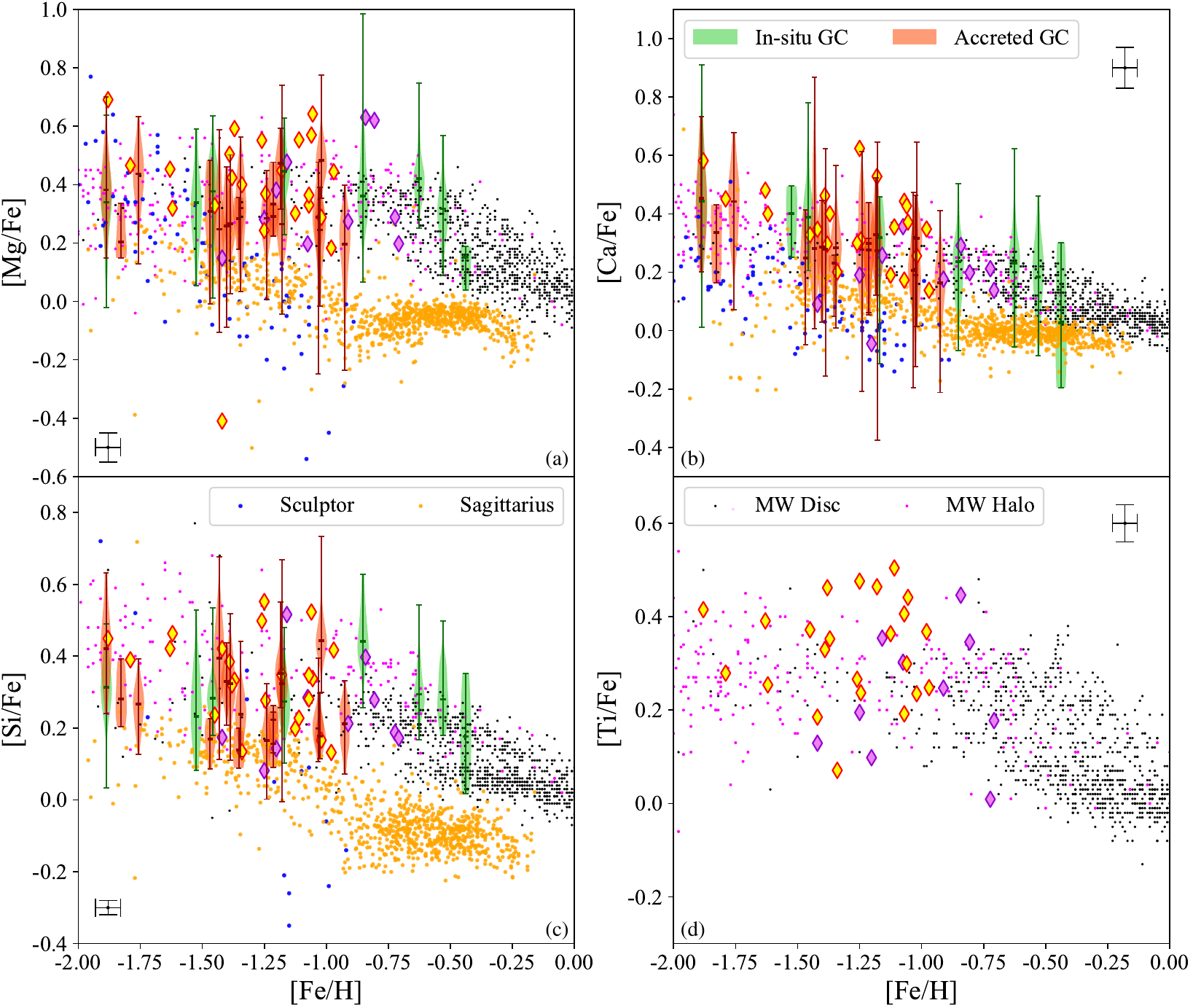}
	\caption{The [$\alpha$/Fe] – [Fe/H] planes, where $\alpha$ elements include Mg, Si, Ca, and Ti. Violin shaped symbols indicate GC stars from APOGEE \citep{Meszaros2020}, including the average [Fe/H], maximum and minimum [$\alpha$/Fe] of each GC. In-situ GCs are colored light green, while accreted GCs are light red. Orange circles correspond to Sagittarius \citep{sgr_hasselquist_2017,sgr-hayes-2020ApJ...889...63H}. Blue circles represent stars from Sculptor \citep{scl_hill2019}. Other symbols are the same as Fig.\ref{fignao}. }
	\label{figalpha}
\end{figure*}

$\hspace*{0.1cm}$C \& N: Since the identification of N-rich field stars from LAMOST low-resolution spectra ($R\sim2000$) using CN and CH molecular bands, there have been concerns about its N-rich nature \citep{tang2019}. Fig.\ref{figcn} clearly shows that the LAMOST-identified N-rich field stars (yellow diamonds) are strongly enhanced in nitrogen ([N/Fe$]> 0.7$ dex). This level of N enhancement is clearly higher than those experienced by normal metal-poor RGB halo stars going through the first dredge-up and internal-mixing (black dots), where obvious C-N anti-correlation is found in the latter sample. Therefore, our LAMOST-identified N-rich field stars cannot be explained by these stellar evolution effects. The [N/Fe] ratios of 10 GALAH/APOGEE N-rich field stars (violet diamonds) are generally lower than those of the LAMOST-identified stars. This is caused by two factors: (1) The [N/Fe] limit between N-rich field stars and normal field stars drawn by \citet{jose_2022A&A...663A.126F} decreases as the metallicity increases. This limit generally reflects the N enhancement by internal-mixing, which is a function of metallicity \citep{lagarde_2012A&A...543A.108L, ShetroneCN}; (2) the GALAH/APOGEE N-rich field stars in this work are more metal-rich, where half of them show [Fe/H] \textgreater $ $ \text{-}0.9 dex. Note that we did not apply the same metallicity cut as the LAMOST-identified stars, because we also want to check the properties of N-rich field stars in higher metallicity region. 

$\hspace*{0.1cm}$O \& Na: GC stars exhibit clear Na-O anti-correlation (background dots in Fig.\ref{fignao}a), indicating the activation of high-temperature proton-capture chains (i.e., ON cycles and NeNa cycles, \citealt{Arnould1999}). Most of the second-generation (SG) stars show [Na/Fe$]>0.15$. In this context, the fact that most N-rich field stars also show [Na/Fe$]>0.15$ agrees with their GC SG escapee scenario. However, N-rich field stars show systematically higher O abundances compared to GC SG stars. When combined with other normal halo field stars, we can see a Na-O anti-correlation for our N-rich field stars (the average [O/Fe] is 0.11 dex lower in the stars with [Na/Fe] $>$ 0.2 compared to those with [Na/Fe] $<$ 0.2). 

$\hspace*{0.1cm}$Mg \& Al: The existence of Mg-Al anti-correlation in GCs indicates the activation of MgAl cycle (Fig.\ref{fignao}b). Based on the FG/SG star definition by their locations on the Na-O plane,  part of the SG stars are not enriched in [Al/Fe], which is comparable to other FG stars and MW field stars. This is caused by the absence of Al enhancement in  metal-rich GC stars \citep{Panico2017,Meszaros2020}. Meanwhile, about half of the N-rich field stars show [Al/Fe$]<+0.5$ dex, similar to the Al-normal SG stars and MW field stars. The overall distribution of our N-rich sample shows close resemblance to that of GC SG sample, again supporting their GC SG escapee scenario.

$\hspace*{0.1cm}$$\alpha$-elements: GC MPs affect not only Mg, but also Si among $\alpha$-elements, particularly in massive or metal-poor GCs \citep{2009A&A...505..139C, Tang2018}. Apart from GC enrichment, the [$\alpha$/Fe]-[Fe/H] planes generally reflect the enrichment history of different galaxies (Fig.\ref{figalpha}), where SNe Ia start at lower metallities, causing lower [$\alpha$/Fe] ratios in dwarf galaxies, e.g, Sculptor (Scl) and Sagittarius (Sgr) dwarf galaxies. Thus, the alpha-to-iron abundance ratios may help decipher the galaxy origins of N-rich field stars (see Section \ref{sec:cd}). 

\begin{figure}[htbp]
	\raggedright
	\includegraphics[width=8.7cm,height=6.75cm]{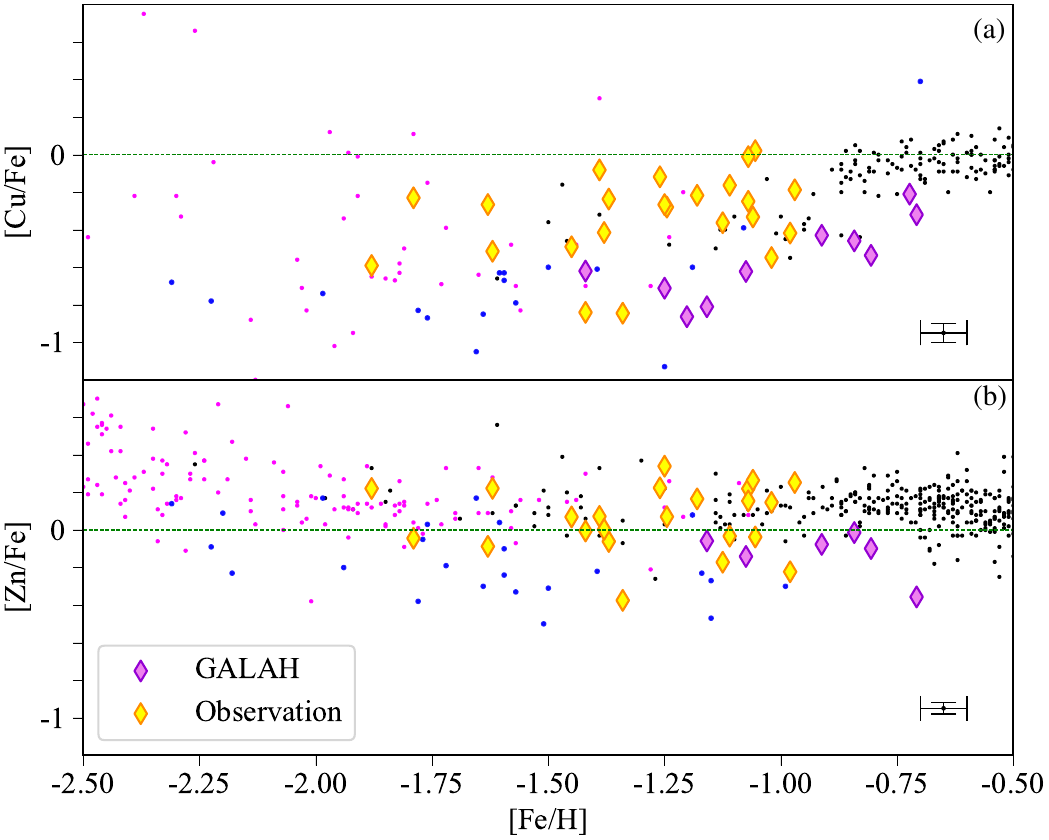}
	\caption{The [Cu/Fe] and [Zn/Fe] vs. [Fe/H] planes.  Blue circles correspond to stars from dwarf galaxies (Sagittarius, Sculptor, Fornax, Carina and Leo I) \citep{sgr_hasselquist_2017,scl_hill2019,DGs_Shetrone2003}. Other symbols are the same as Fig.\ref{figalpha}. }
	\label{figcuzn}
\end{figure}

$\hspace*{0.1cm}$Cu \& Zn: Cu is one of the odd-Z elements with metallicity-dependent yields, thus [Cu/Fe] show an increasing trend toward higher metallicities (Fig.\ref{figcuzn}a), similar to the galactic evolution of Na and Al \citep{ele_ori_2020}. However, Cu does not participate in the high-temperature hydrogen burning, therefore we do not expect variations in this element. 
Meanwhile, Zn is produced mainly through explosive Si burning in core-collapse supernovae. However, the increasing [Zn/Fe] of halo stars (magenta dots) toward lower metallicities (Fig.\ref{figcuzn}b) indicates that a large fraction of hypernovae is needed in this region \citep{Nomoto2013}. The N-rich field stars generally follow the [Cu/Fe] (and [Zn/Fe]) trend of other field stars.  Interestingly, [Zn/Fe] abundances of dwarf galaxies are systematically lower compared to other MW field stars \citep{skuladottir-scl-2017A&A...606A..71S} in the metallicity range of [$-2.0, -1.0$], which helps to discriminate their progenitor galaxies in Section \ref{sec:cd}. 

$\hspace*{0.2cm}$In summary, our N-rich field stars exhibit O, Na, Mg, and Al abundance distributions similar to those of GC SG stars. Meanwhile, their Si and Ca abundances fall within the typical GC ranges. This overall pattern is consistent with the GC escapee scenario. We did not further compare chemical abundances of the iron-peak elements and neutron-capture elements between N-rich field stars and GC stars, since most GC stars do not show variation in these elements (except a few Type II GCs). GC stars generally show similar chemical abundances in these elements compared to MW field stars. And so do N-rich field stars \citep{yu2021}. 

$\hspace*{0.2cm}$The obvious scatters in light and $\alpha$ elements, partial chemical kinship with dwarf galaxies yet kinematic assimilation into the halo, indicate these stars are likely to originate from different globular clusters either accreted from dwarf satellites or formed in the Milky Way. To test whether these chemical groups trace distinct progenitor systems, we performed orbital analysis in the next subsection. 

\subsection{Chemodynamical grouping}
\label{sec:cd}

$\hspace*{0.2cm}$Since the GC origins of N-rich field stars are supported by their light and $\alpha$ elements, is it possible to identify their host GCs as in-situ or accreted? Given that $\sim 80$\% of GC escapees are lost to the field through tidal evaporation \citep{Weatherford2023}, most of the N-rich field stars should maintain similar kinematics as their host GCs. Moreover, the heavier element abundances are likely to be homogeneous among GCs, except for a few type II clusters \citep{Milone2017}. Assuming similar chemodynamical features between N-rich field stars, host GCs, and progenitor galaxies \citep{Massari2019,LinSH2025}, studying the dynamics and chemistry of N-rich field stars may help us to obtain information about their origins. 

$\hspace*{0.2cm}$In this work, the orbital parameters (including orbital energy E, angular momentum L, eccentricity e, inclination i) of all stars were calculated with the publicly available Python library GALPY \citep{2015ApJS..216...29B}. Specifically, e$=\frac{r_{apo}-r_{peri}}{r_{apo}+r_{peri}}$, and i = arccos($L_z/L$), where $r_{apo}$ and $r_{peri}$ are the apocenter radius and pericenter radius, respectively. We adopted the steady-state and axisymmetric builtin MW potential: MWpotential2014\footnote{Switching to newer MW potential, like the one in \cite{m17_2017MNRAS.465...76M}, does not significantly change our results. }. The values of the solar Galactocentric radius and the local circular velocity are set to \textit{$R_\odot$ = 8 kpc}, \textit{$v_\odot$ = 220 km $s^{-1}$}, respectively \citep{bovy_2012ApJ...759..131B}. The peculiar velocity ([$U_\odot$, $V_\odot$, $W_\odot$]) of the Sun is set to [11.1, 12.24, 7.25] km $s^{-1}$ \citep{2010MNRAS.403.1829S}. The adopted distances of 23 LAMOST-identified N-rich stars are Bayesian spectrophotometric
distances with no assumptions about the underlying populations \citep{tang2020}, while STARHORSE \citep{starhorse2018} distances are adpoted for the GALAH/APOGEE stars. We also analyzed the orbital parameters of more than five thousand MW stars selected from the GALAH DR4 catalog crossmatched with APOGEE DR17 and limited the sample to stars with a radial velocity error (from APOGEE) smaller than 0.2 km $s^{-1}$ and $\mathrm{RV_{scatter}}$ $<$ 1 km $s^{-1}$ (the main purpose for this selection is to ensure the accuracy of astrometry and the reliability of subsequent Galpy orbital analysis). 

\begin{figure}[htbp]
	\raggedleft
	\includegraphics[scale=0.268]{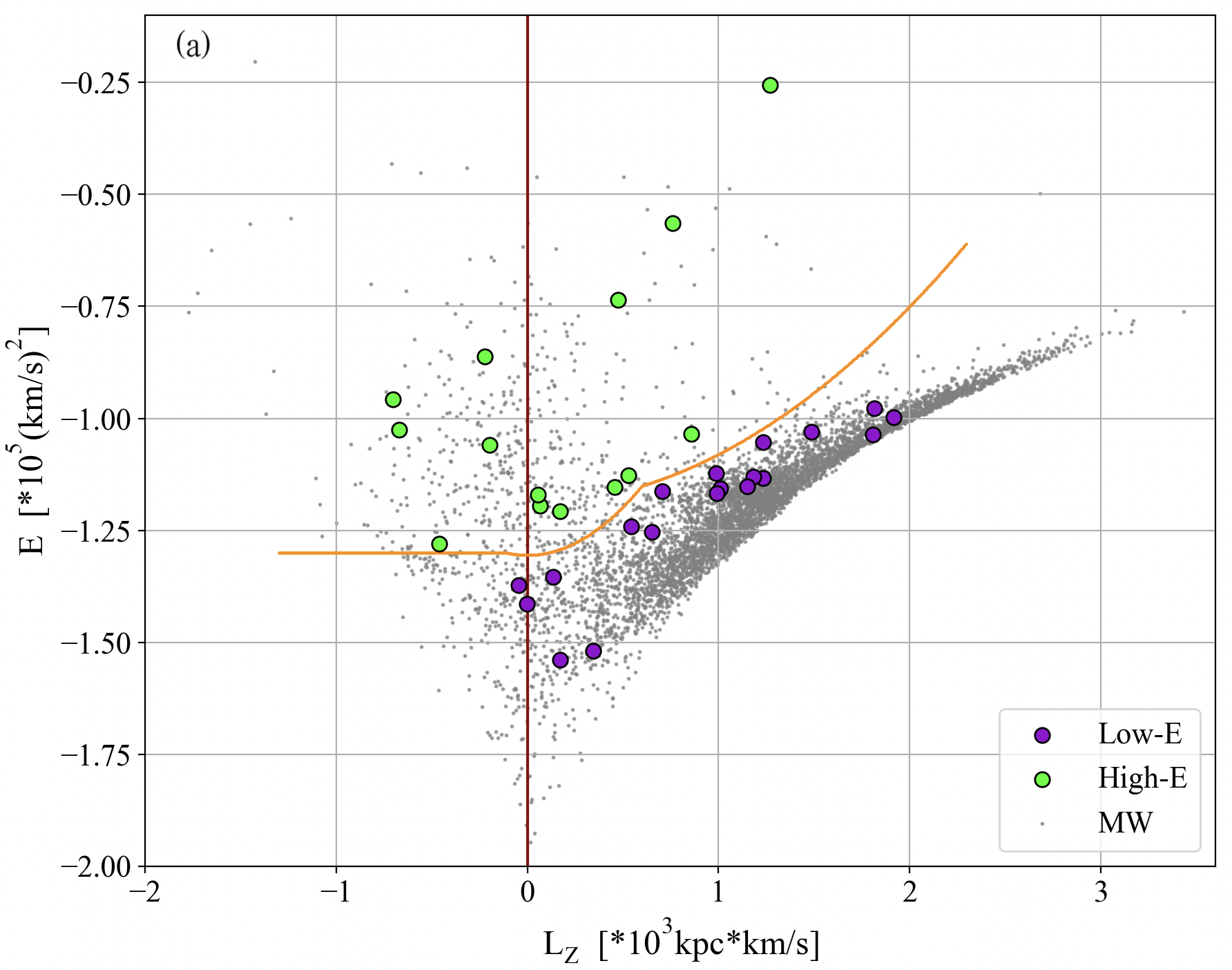}
	\includegraphics[scale=0.268]{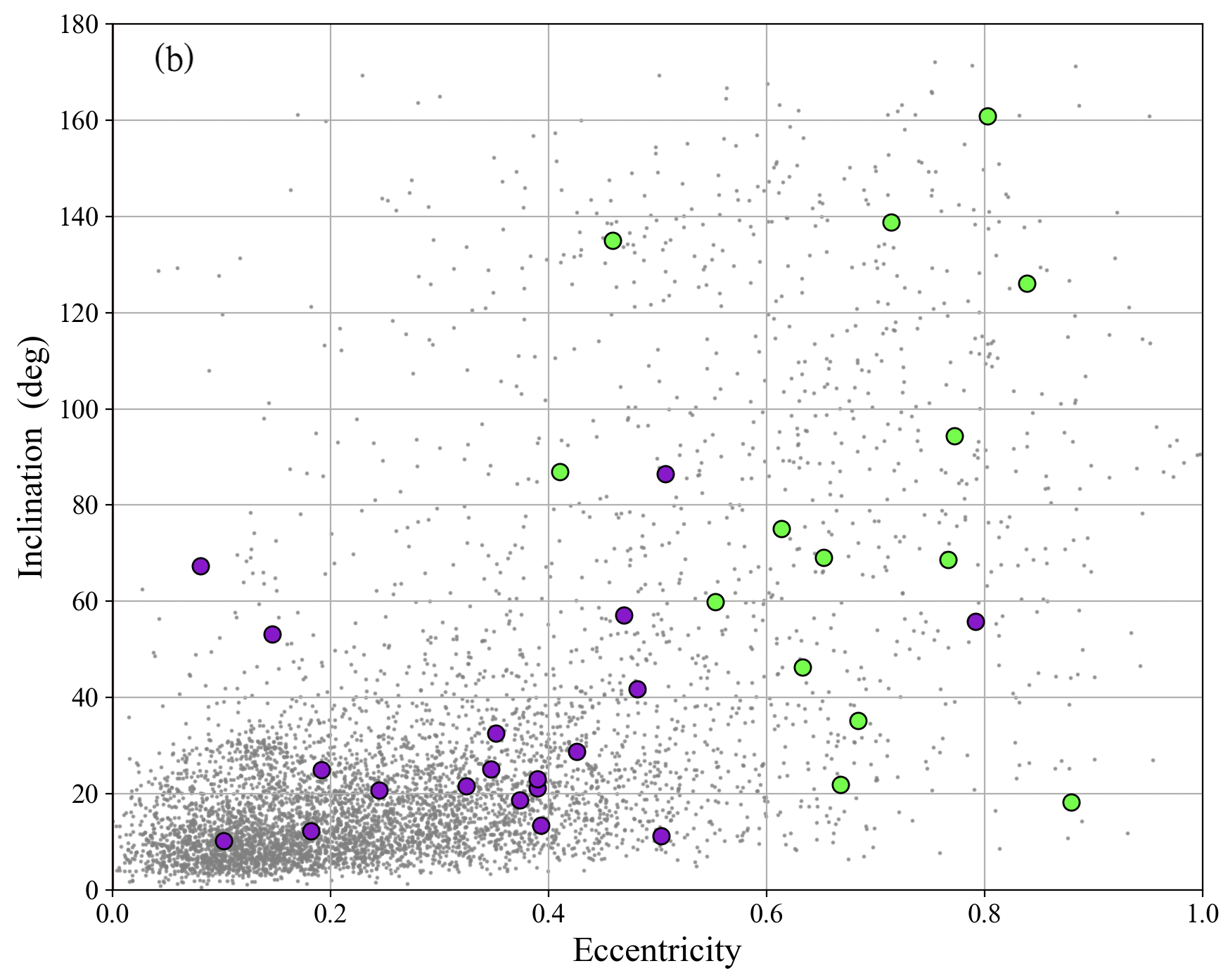}
	\caption{Panel a: The diagram of E \textit{vs} $L_Z$ of N-rich stars (circles) in two groups and GALAH MW stars (gray dots). HE stars are colored lime and LE stars are dark violet, and the dark orange line shows the boundary between them. See text for more details about their classification. Panel b: The stars' distribution in orbital inclination and eccentricity. }
	\label{figdyn}
\end{figure}

\begin{figure*}[htbp]
	\raggedright
	\includegraphics[width=17cm,height=10.5cm]{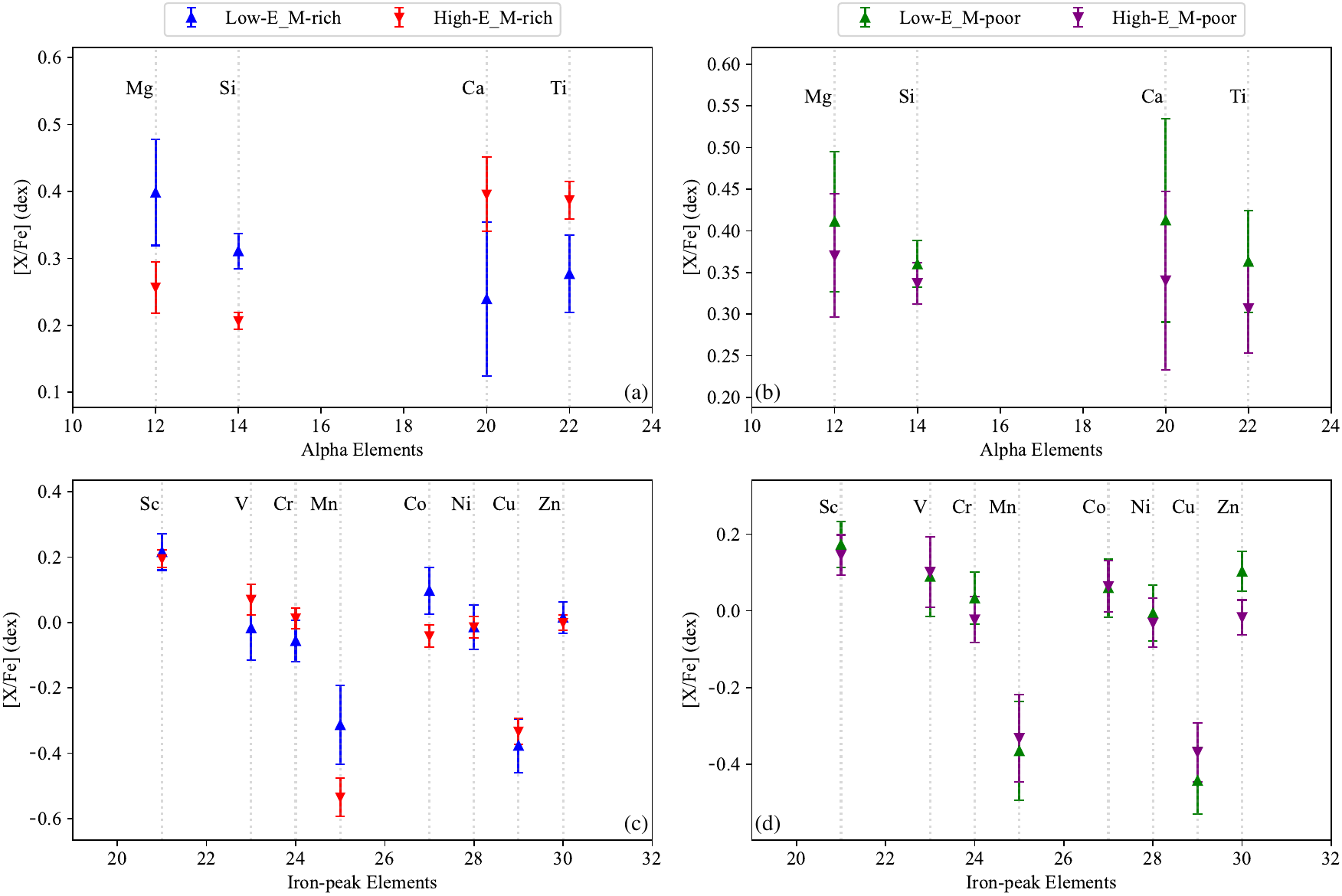}
	\caption{The chemical patterns of different groups, including $\alpha$ elements, iron-peak elements, Cu and Zn. Triangles stand for the average element abundances of 4 groups with corresponding errors (blue for LE-MR, red for HE-MR, green for LE-MP and purple for HE-MP). }
	\label{fig2group}
\end{figure*}

\begin{figure}[htbp]
	\raggedright
	\includegraphics[width=8.9cm,height=9.1cm]{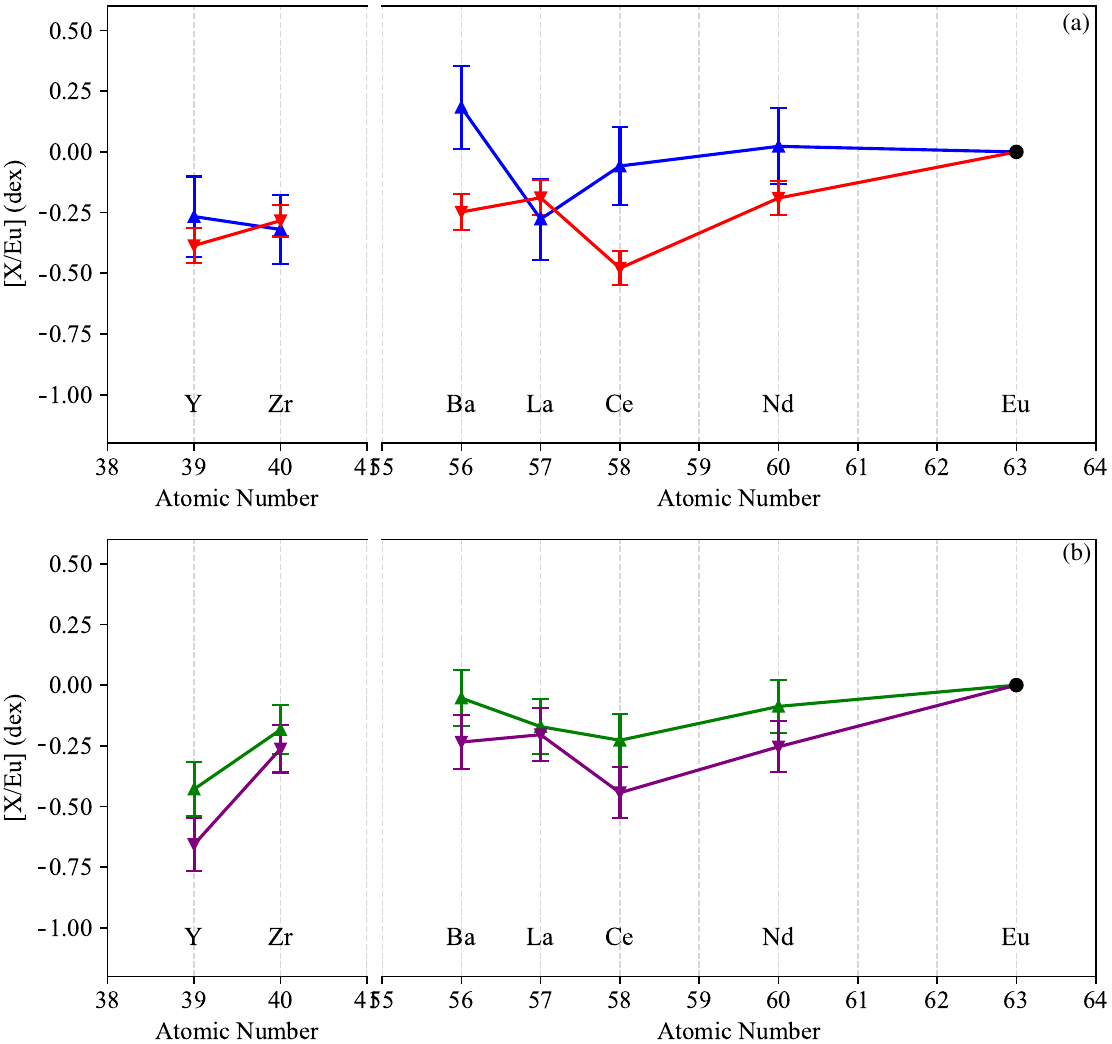}
	\caption{The chemical patterns of neutron-capture elements of different groups in the form of [X/Eu]. Markers are the same as Fig.\ref{fig2group}. }
	\label{figxeu}
\end{figure}

\begin{figure}[htbp]
	\raggedright
	\includegraphics[width=8.8cm,height=5.2cm]{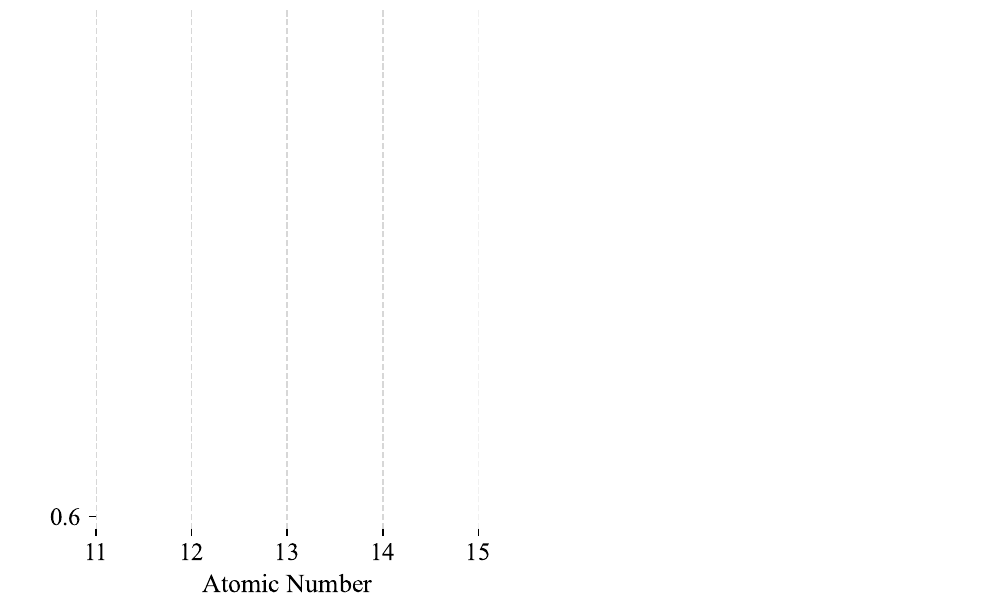}
	\caption{[Mg/Eu] and [Si/Eu] of different groups. Markers are the same as Fig.\ref{fig2group}. }
	\label{figalpeu}
\end{figure}

$\hspace*{0.2cm}$ We divided our sample stars into two groups (Fig.\ref{figdyn}) according to their positions in E-$L_z$ phase space\footnote{The boundary is based on that of \citet{belo24_2024MNRAS.528.3198B}. }. (1) High-Energy group (High-E group, or HE): These stars generally exhibit higher orbital energies (E) and high eccentricities. Their dynamics align with known accreted structures like the Gaia-Sausage-Enceladus \citep[GSE,][]{igs_2021MNRAS.500.1385H}. (2) Low-Energy group (Low-E group, or LE): These stars generally show lower orbital energies, mostly positive $L_z$, smaller inclination angles, and lower eccentricities. Their dynamical similarity to the MW thick disk stars implies in-situ origins. In order to compare the chemical abundances of these two groups, it is optimal to use an overlapping metallicity range. After balancing the bin sizes and sample sizes, we divided each group into two with [Fe/H]$ = -1.1 $ as the boundary, so there are 4 subgroups: high-energy metal-poor (HE-MP), low-energy metal-poor (LE-MP), high-energy metal-rich (HE-MR), and low-energy metal-rich (LE-MR). Their mean metallicities are $-1.38\pm0.186$, $-1.38\pm0.232$, $-1.02\pm0.039$, $-0.94\pm0.134$, respectively. Such partitioning allows targeted investigation of how birth GC environments (accreted vs. in-situ) imprint distinct chemical signatures on their escaped stars. The systematic abundance differences between high-energy (HE) and low-energy (LE) N-rich stars (Figs.\ref{fig2group} \& \ref{figxeu} \& \ref{figalpeu}) provide critical insights into their distinct progenitor environments and Galactic assembly history. 

$\hspace*{0.1cm}$$\alpha$ elements: As we mentioned earlier in Section \ref{sec:abu}, stars born in dwarf galaxies generally show lower [$\alpha$/Fe] compared to in-situ stars at [Fe/H] $\gtrapprox -1.5$. Assuming HE stars are mostly accreted, while LE stars are mostly in-situ, we expect lower [$\alpha$/Fe] for HE stars compared to LE stars. This agrees with the observations in Fig.\ref{fig2group}, except for Ca and Ti in the metal-rich region. Such exception may be caused by two reasons: (1) Accreted (especially from GSE) and in-situ stars show smaller differences in [Ca/Fe] and [Ti/Fe] at a given metallicity \citep{2021ApJ...923..172H}; (2) HE-MR stars are on average more metal-poor than LE-MR stars ($-1.02\pm0.039$ versus $-0.94\pm0.134$). We also note that HE and LE stars show smaller [$\alpha$/Fe] differences for MP groups, since in-situ stars and accreted stars show similar [$\alpha$/Fe] at [Fe/H] $< -1.5$, and the abundances of Mg and Si are possibly affected more obviously by H-burning in metal-poor GC environments. 

$\hspace*{0.1cm}$Iron-peak elements: Sc, V, Cr, Mn, Co, and Ni are substantially generated by SNe Ia, thus showing close correlations with Fe. HE and LE stars do not show significant differences in these elements, except for Mn and Co in the metal-rich regime. Such exception may be caused by the lower production of Mn and Co in SNe Ia of dwarf galaxies \citep{Sanders2021,yu2021}. Contrary to our results, \cite{NS10} and \cite{fishlock_2017MNRAS.466.4672F} suggested that in-situ and accreted populations show nontrivial differences in [Ni/Fe] and [Sc/Fe]. Admittedly, the lower spectral resolution and lower SNR in our work compared to theirs result in lower precision (0.1 dex vs. 0.01 dex), possibly causing the non-detection of Sc and Ni differences between the two sub-populations. 

$\hspace*{0.1cm}$Cu \& Zn: In the primordial MW where rapid star formation and frequent Type II SNe enhanced Zn production via $\alpha$-rich freeze-out, while earlier SN Ia contributions suppressed Zn enrichment in dwarf galaxies. On the other hand, Cu in nearby dwarf spheroidal galaxies show similar abundances as MW halo stars \citep{DGs_Shetrone2003}. In this work, we find that HE and LE groups show similar [Cu/Fe] values within the uncertainty ranges. Different from Cu, LE stars exhibit higher [Zn/Fe] than HE stars in MP groups ($\Delta$ $\sim$ 0.15 dex), consistent with the in-situ formation feature. Conversely, HE stars’ lower [Zn/Fe] agree with other dwarf galaxies \citep{skuladottir-scl-2017A&A...606A..71S,DGs_Shetrone2003}. 

$\hspace*{0.1cm}$Neutron-capture elements: The systematic differences in neutron-capture element abundances ([X/Eu]) between high-energy (HE) and low-energy (LE) nitrogen-rich stars are illustrated in Fig.\ref{figxeu} and provide specific clues to their origins and the Milky Way’s assembly history. HE stars exhibit lower [Y/Eu], [Zr/Eu], and [Ba/Eu] ratios compared to LE stars, reflecting a dominance of r-process nucleosynthesis over the s-process. Besides the [X/Eu] ratios displayed, HE stars also show a higher mean [La/Zr] ratio and lower mean [Zr/Fe], consistent with the chemical pattern of low-$\alpha$ sample studied in \cite{fishlock_2017MNRAS.466.4672F}. This pattern arises because HE stars likely formed in massive dwarf galaxies (e.g., the Gaia-Sausage-Enceladus remnant), where core-collapse supernovae and delayed neutron star mergers significantly contributed to the nearly pure {\it r}-process enrichment \citep{Sneden2008_n-capture,2022gse_rpro}. In contrast, LE stars, presumably associated with the in-situ populations, show enhanced [X/Eu], [Zr/Fe] and lower [La/Zr] due to prolonged contributions from AGB stars, which efficiently produce {\it s}-process elements \citep{mw_agb_spro_2014,ele_ori_2020}. Moreover, we find that the LE stars exhibit systematically higher mean [$\alpha$/Eu] (using Mg and Si as $\alpha$-element representatives) than the HE stars across both metal-poor and metal-rich regimes (see Fig.\ref{figalpeu}). This pattern aligns with established results where accreted stars (like GSE remnants) show elevated [Eu/$\alpha$] compared to MW in-situ stars \citep{matsuno_2021A&A...650A.110M,monty_2024MNRAS.533.2420M}. On the whole, the persistence of low [X/Eu] and other features mentioned in HE groups support our chemodynamical inference, underscoring their distinct accretion origin independent of in-situ galactic chemical evolution. These findings align with hierarchical galaxy formation models, where the MW incorporated chemically distinct populations via mergers, as evidenced by dynamical studies linking HE kinematics to disrupted massive satellites \citep{Ji2016,helmi_Natur.563...85H}. 

$\hspace*{0.2cm}$In conclusion, the observed discrepancies between HE and LE stars collectively suggest a dichotomy in their chemodynamical properties, pointing to distinct progenitor environments. HE stars might have been escaped from GCs belonging to disrupted dwarf galaxies, where the enrichment processes were less efficient or occurred over a longer timescale. Conversely, LE stars, which are more likely to be GC escapees confined to the MW disk reflecting a history of sustained chemical enrichment. 

\section{Discussion}
\label{sec:4}

\subsection{Connection between N-rich field stars and high-redshift N-emitters}

\begin{figure}[htbp]
	\raggedright
	\includegraphics[width=8.6cm,height=7.1cm]{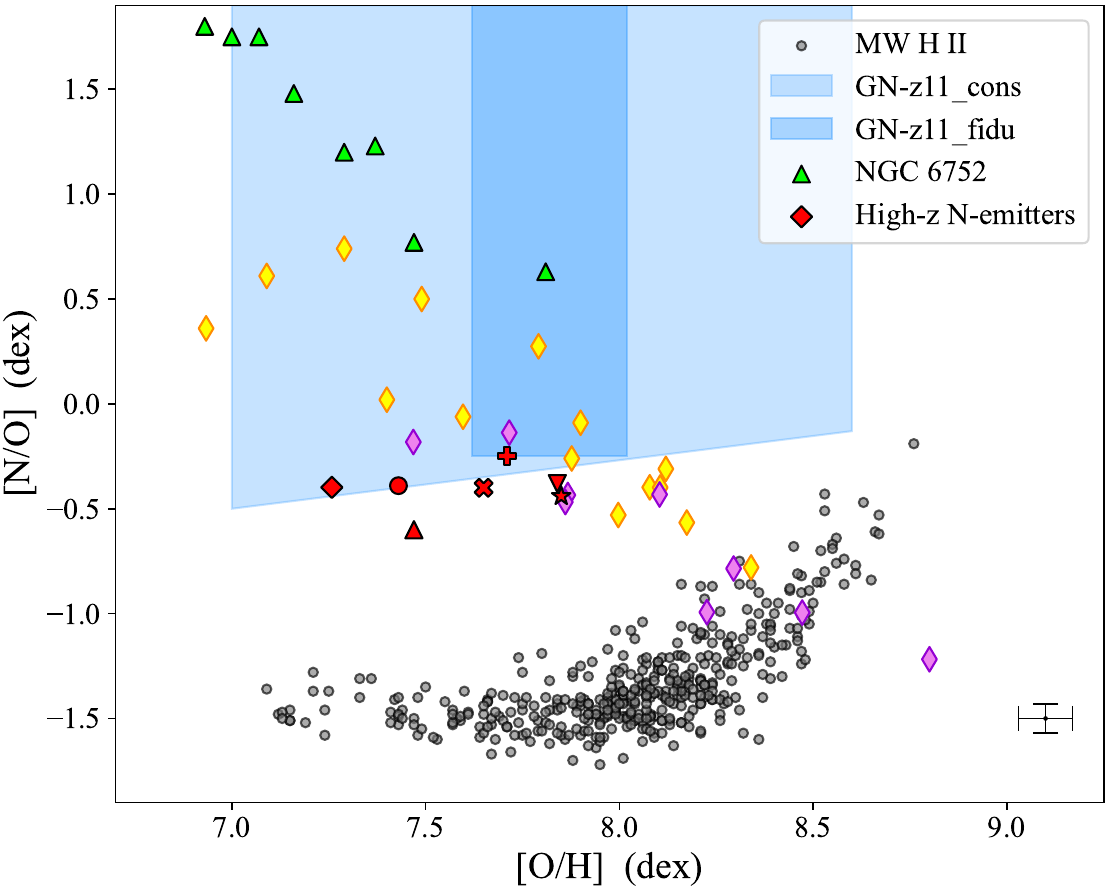}
	\caption{The [N/O] - [O/H] plane. Small gray circles show the Galactic H II region abundances from \cite{h2_2012MNRAS.421.1624P}. The conservative (fiducial) range of [N/O] and [O/H] inferred for GN-z11 from \cite{gn-z11-fiducial} is displayed as the light (dark) dodgerblue shaded region. Values for high-z N-emitters \citep{gn-z11-dots,highz-gnz11-2024arXiv240714201N,highz-ghz2-2017MNRAS.472.2608S,glass-isobe2023,topping2024,Topping2025} are shown as red markers and lime triangles are measurements from stars in NGC 6752 \citep{ngc6752-carretta}. }
	\label{figno}
\end{figure}

$\hspace*{0.2cm}$After the launch of JWST, astronomers are reevaluating the scientific significance of GCs in the high-redshift Universe. Observations now indicate their dominant role on galaxy evolution at these redshifts. Clumpy and bursty star formation on parsec scale observed at high redshift ($z \sim 8-11$) suggests that these are proto-GC actively forming \citep[e.g.,][]{PROTO-GC-2024A&A...681A..30M}. Furthermore, newly discovered ``N-emitters'' exhibiting [N/O] ratios similar to stars in local GCs, indicating they might be also forming GCs \citep[e.g.,][]{topping2024,Topping2025}. 

$\hspace*{0.2cm}$In this work, we show that N-rich field stars are possible candidates of GC escaped stars. Do they show similar [N/O] ratios as those high-redshift N-emitters or local GCs? To investigate their connections, we compared their [N/O] \textit{vs.} [O/H] ratios in Fig.\ref{figno}, including local H II regions for reference. Our sample stars reveal an anti-correlation between [N/O] and [O/H], a trend also observed in stars within the GC NGC 6752. Most N-rich field stars occupy the same region in this plane as high-redshift N-emitters, implying a similar enrichment history. Exceptions occur among the more metal-rich N-rich field stars ([Fe/H$] \gtrapprox -1.0$), whose [N/O] ratios align more closely to local H II regions. This chemical pattern suggest that N-rich field stars with [Fe/H$] \lessapprox -1.0$ are local analogs of the high-redshift N-emitters. On the contrary, N-rich field stars with [Fe/H$]>-1.0$ may not have similar chemical enrichment mechanism as N-emitters. Consequently, these findings support the validity of the metallicity cut applied to LAMOST-identified N-rich field stars \citep{tang2019,tang2020}. 

\subsection{Tracing origins among globular clusters}

\begin{figure}[htbp]
	\centering
	\includegraphics[width=8.33cm,height=6.25cm]{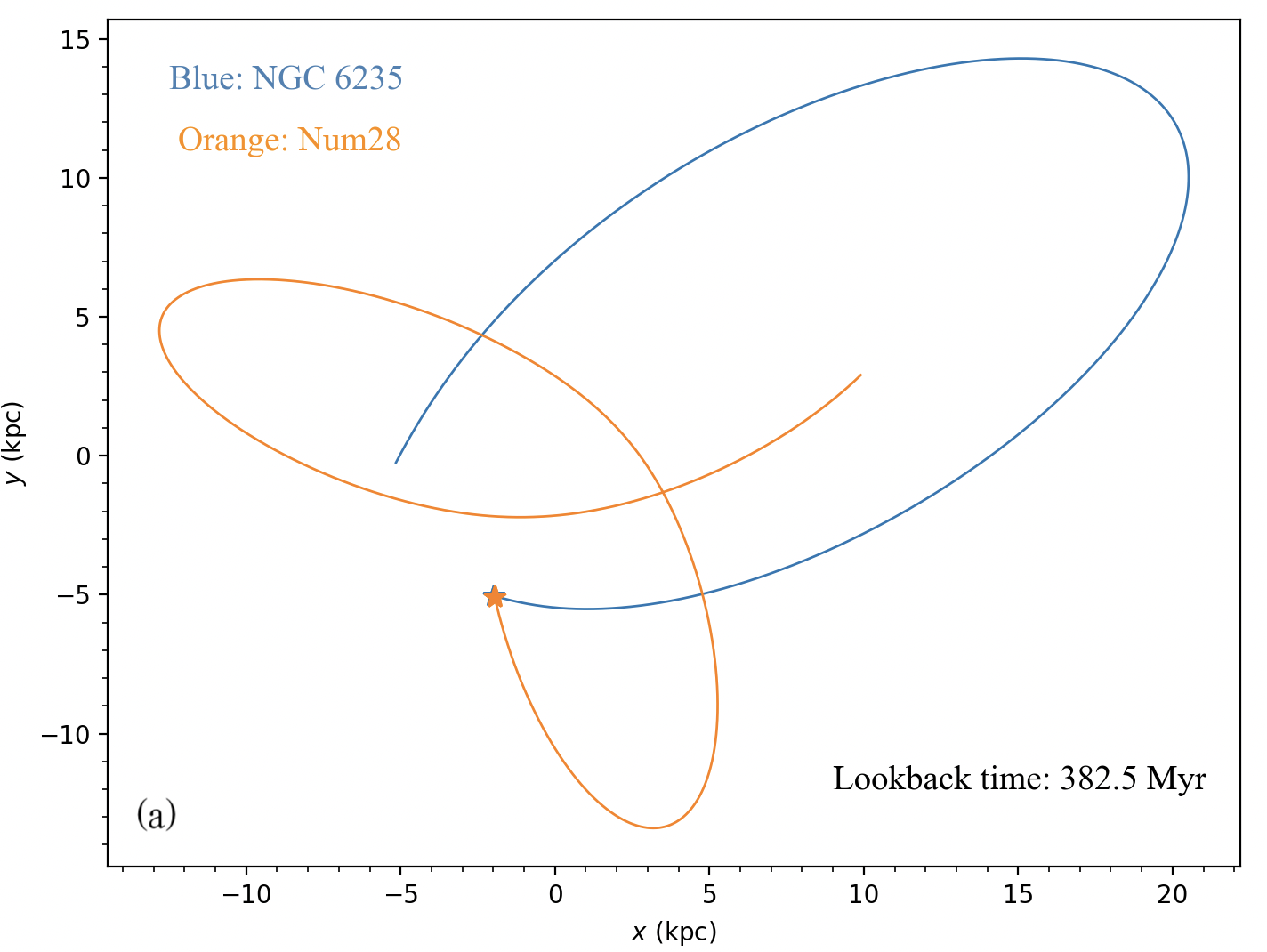}
	\includegraphics[width=8.4cm,height=6.25cm]{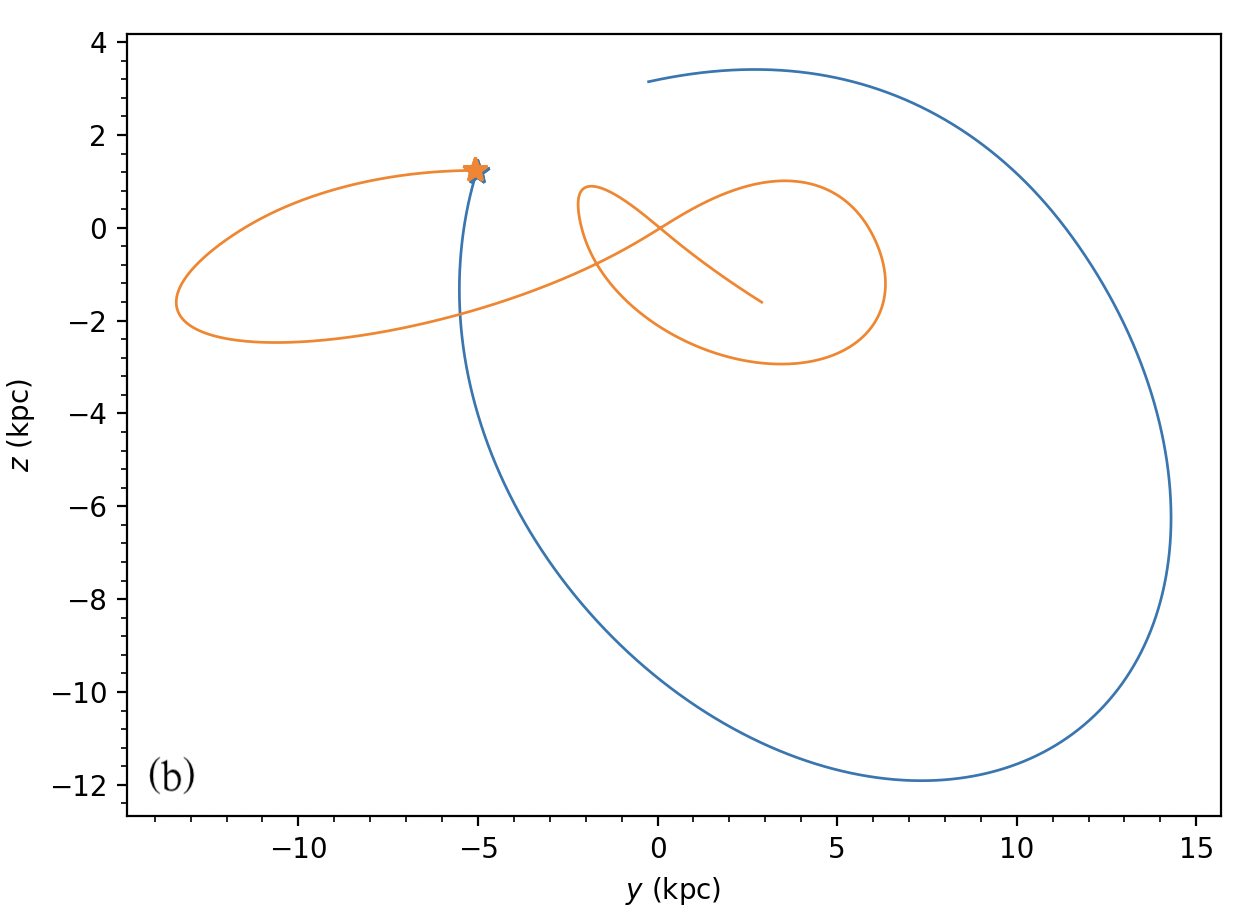}
    \includegraphics[width=8.3cm,height=6.25cm]{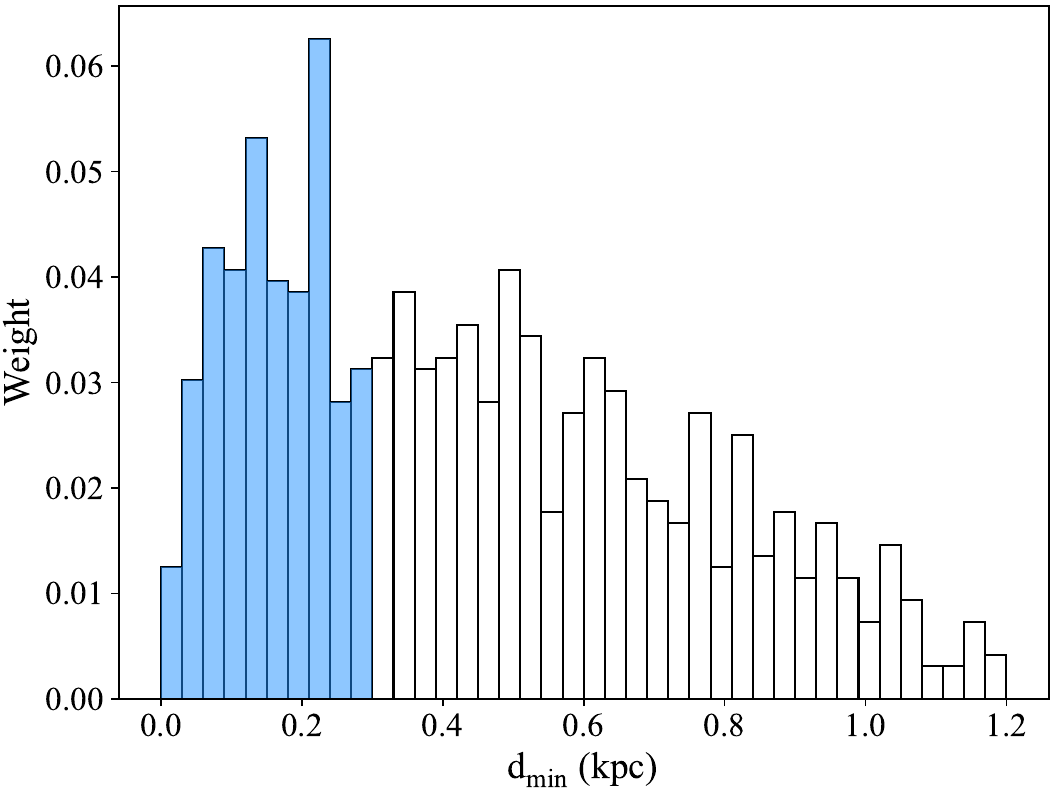}
	\caption{Panel a \& b: Orbital integrations of NGC 6235 and Star Num28. Filled stars stand for their spatial locations at the look-back time and lines indicate their orbits (dodger blue for NGC 6235 and orange for Num28). Panel c: The distribution of minimum distances between the integrated orbits of Num28 and NGC 6235 revealed by the Monte Carlo analysis. }
	\label{figorbit}
\end{figure}

$\hspace*{0.2cm}$The concurrent spatial proximity and chemical coherence between nitrogen-enhanced field stars and specific GCs provide most compelling evidence for dynamical escape scenarios \citep[e.g.,][]{souza_ter5_2024ApJ...977L..33S,HuangY2025}. In this work, we have attempted to provide direct evidence linking the N-rich stars to their possible host GCs. We calculated the orbital elements of more than 100 known GCs using the kinematic parameters reported in \cite{gcs_baum_2019MNRAS.482.5138B}, and for each specific N-rich star we selected candidate GCs which are relatively similar to it in the E-$L_z$ phase space \citep{Xu2024}. Then we performed orbit integrations of both the stars and candidate GCs with Galpy to see if they would get close in space while tracing back. Among all the matching results, the most representative pair is star Num28 and NGC 6235. Orbit integrations revealed a close encounter in space ($<$ 0.1 kpc, comparable to the typical tidal radii of GCs) between them about 380 Myrs ago (panel a \& b of Fig.\ref{figorbit}). To explore the effect of orbital uncertainties, we applied a Monte Carlo analysis of the orbit integration. We performed 1000 orbit integrations by randomly perturbing the proper motions and distances of Num28 and NGC 6235. These parameters were drawn from normal distributions with ${\sigma_1}$ = 0.02 mas $\mathrm{yr^{-1}}$ for $\mathrm{\mu_{\alpha*}}$ and $\mathrm{\mu_{\delta}}$, ${\sigma_2}$ = 0.2 kpc for distance around the measured values. The results revealed that in about 36\% of the simulations, the minimum separation distance between Num28 and NGC 6235 during the integration was smaller than 0.3 kpc (panel c of Fig.\ref{figorbit}). 

$\hspace*{0.2cm}$This evidence about orbits encounter is part of a combined chemodynamical argument, which includes not only kinematic consistency in E–Lz space but also chemical similarity. The [Fe/H] of star Num28 matches that of NGC 6235 ($\sim -1.1$ dex) \citep{6235_2003AJ....125..801H} within 0.1 dex and the light-element abundances indicate that Star Num28 is an SG star with enhanced N, Na, and Al ([N/Fe]$= 1.09$, [Na/Fe]$= 0.71$, [Al/Fe]$= 0.53$). The close encounter between star Num28 and NGC 6235, is intriguing given their high relative velocity ($>$ 300 km $s^{-1}$). This high speed is difficult to explain through standard tidal stripping alone and suggests a more violent ejection mechanism. One probable scenario is that Num28 experienced a gravitational ``kick'' from an intermediate-mass black hole within the cluster, which would impart the high velocity required for escape, providing a dynamic alternative to the gradual stripping of stars \citep{HuangY2025}. These combined chemodynamical analysis suggest that star Num28 is a likely SG escapee from NGC 6235, contributing to our understanding of globular cluster dissolution and the Galactic stellar halo’s origins. 

$\hspace*{0.2cm}$Meanwhile, uncertainties persist. The orbital encounter rate reflects the sensitivity of long-term orbital integration to distance uncertainties, therefore, improving distance precision is a key to enable more robust investigations in the future. Also, assumptions in Galactic potential models and others may influence orbit reconstruction. Future studies should prioritize high-precision astrometry, such as upcoming Gaia data releases, to test the robustness of the association. By transforming N-rich stars from chemical curiosities into quantitative probes, this work may provide a template for reconstructing the Milky Way’s hierarchical assembly --- one escaped star at a time. 

\section{Summary}
\label{sec:5}

$\hspace*{0.2cm}$This study establishes nitrogen-enhanced field stars as definitive tracers of globular cluster (GC) disruption and Galactic accretion by synthesizing orbital dynamics and multi-element chemical abundances. This work is part of our ongoing research project, ``Scrutinizing GAlaxy-STaR cluster coevolutiON with chemOdynaMIcs (GASTRONOMI)'', which leverages multi-wavelength photometric and spectroscopic data to unravel the coevolutionary relationships between the MW, its satellite dwarf galaxies, and their star clusters. Using high-resolution spectroscopy of ESPaDOnS, MIKE and GALAH for 33 N-rich stars, we analyzed their chemodynamical features and confirmed evident chemical similarity in multi-element abundances between these stars and the GC second generation stars, reinforcing the association between N-rich field stars and globular clusters. 

$\hspace*{0.2cm}$To explore the possible formation pathways, we classified the stars into different groups according to orbital energy and metallicity. HE groups exhibit orbital parameters aligning with accreted massive dwarf galaxies, like GSE, meanwhile, LE groups show disk-like kinematics consistent with in-situ populations. And their distinct chemical patterns suggest different progenitor environments: HE groups trace clusters stripped from massive dwarf galaxies with enhanced r/s-process ratios, and LE groups are likely to originate from GCs formed within the Milky Way, enhanced in $\alpha$ elements relative to their counterparts formed in dwarf galaxies and accreted into MW. 

$\hspace*{0.2cm}$We also found a chemical connection between metal-poor Milky Way N-rich field stars and high-redshift "N-emitters." They occupy the same region in the [N/O] vs. [O/H] plane, suggesting a similar enrichment history and indicating that the local N-rich stars are analogs of the high-z systems, possibly linked to globular cluster formation. Furthermore, using orbital integration, we traced possible second generation stars of peculiar globular clusters and this method can be applied to larger samples with more accurate stellar distances. 

$\hspace*{0.2cm}$These results help resolve long-standing ambiguities in N-rich field stars' origins: their chemodynamical clustering decouples accretion signatures from in-situ disruption. All in all, these cases collectively transform single-star archaeology into a tool for reconstructing cluster dissolution histories. 

$\hspace*{0.2cm}$There are still critical avenues demanding exploration in the future. Current samples lack metal-poor stars ([Fe/H] $<-2.0$ dex), where pristine accretion signatures may dominate. High-resolution UV spectroscopy is essential to measure the abundances of CNO and {\it r}-process elements in this metallicity regime. Measurements of CNO are crucial as they directly probe the proton-capture processes unique to the multiple populations within GCs. Neutron-capture elements, especially {\it r}-process elements which are more difficult to measure, serve as tracers to distinguish whether the ancestral systems formed in-situ or were accreted with distinct chemical evolution histories. Another problem is the limited multi-element data, especially for the neutron-capture elements. Future surveys (e.g., 4MOST, WEAVE) will provide more Eu/Ba/La/Ce measurements for chemical peculiar stars to map {\it r/s}-process cartography in accreted systems. There is no doubt that larger samples available and more comprehensive chemical abundance data will provide crucial assistance for the studies of Galactic archaeology with chemical peculiar stars. 

\begin{acknowledgements}
	$\hspace*{0.4cm}$Y. Qiao, B. Tang gratefully acknowledge support from the National Natural Science Foundation of China through grants NOs. 12233013 and 12473035, China Manned Space Project under grant NO. CMS-CSST-2025-A13 and CMS-CSST-2021-A08, the Fundamental Research Funds for the Central Universities, Sun Yat-sen University (24qnpy121). J.G.F-T gratefully acknowledges the grants support provided by ANID Fondecyt Postdoc No. 3230001 (Sponsoring researcher), the Joint Committee ESO-Government of Chile under the agreement 2023 ORP 062/2023, and the support of the Doctoral Program in Artificial Intelligence, DISC-UCN. C. Allende Prieto acknowledges financial support from the Spanish Ministry of Science, Innovation and Universities (MICIU) projects PID2020-117493GB-I00, PID2023-149982NB-I00 and PID2023-146453NB-I00 (\textit{PLAtoSOnG}). B. Barbuy acknowledges partial financial support from FAPESP, CNPq, and CAPES. T. Masseron acknowledge support from the Spanish Ministry of Science and Innovation with the grant No. PID2023-146453NB-100 (\textit{PLAtoSOnG}). Z. Y. is supported by the National Key R\&D Program of China via 2024YFA1611601. 
\end{acknowledgements}

\software{TOPCAT \citep{taylor2005}, BACCHUS \citep{bacchus_2016}, iSpec \citep{ispec_2018JChEd..95...97V}, PySME \citep{pysme_2021csss.confE...1W}, Matplotlib \citep{Hunter2007}, Numpy \citep{numpy2011,numpy2020}, Math, Pandas \citep{pandas2010}, Astropy \citep{astropy2013,astropy2018}, Scipy \citep{scipy2020}, Seaborn \citep{seaborn2021}, Telluric-Filter, Galpy \citep{2015ApJS..216...29B}, StarHorse \citep{starhorse2018}. }

\bibliography{sample}{}
\bibliographystyle{aasjournal}

\appendix
\setcounter{table}{0}
\renewcommand{\thetable}{A\arabic{table}}
\setcounter{figure}{0}
\renewcommand{\thefigure}{A\arabic{figure}}

\begin{table}[htbp]
	\centering
	\caption{The derived chemical abundances of 10 N-rich field stars with GALAH observations are listed here. The typical errors for all elements are also shown. }
	\label{tab:a1}
	\begin{tabular}{c c c c c c c c c c c c}
		\hline\hline\noalign{\smallskip}	
		Elements & G1 & G2 & G3 & G4 & G5 & G6 & G7 & G8 & G9 & G10 & Error$^a$ \\
		\noalign{\smallskip}\hline\noalign{\smallskip}
		RA (J2000) & 289.4999 & 229.9611 & 330.4666 & 342.1356 & 91.8436 & 357.8821 & 330.2343 & 243.0392 & 73.4694 & 243.0330 & - \\
		Dec (J2000) & -29.3267 & 2.3648 & -11.6966 & -23.8799 & -58.8617 & -3.2086 & -8.1948 & -24.6645 & -64.1526 & -25.0677 & - \\
		{[Fe/H]} & -1.202 & -0.723 & -1.420 & -1.159 & -0.806 & -0.708 & -0.912 & -1.075 & -0.842 & -1.250 & 0.05 \\
		{[C/Fe]} & -0.37 & -0.23 & -0.48 & 0.00 & 0.04 & 0.03 & 0.03 & 0.08 & 0.08 & -0.34 & 0.04 \\
		{[N/Fe]} & 0.86 & 0.53 & 0.93 & 0.61 & 0.54 & 0.62 & 0.80 & 0.87 & 0.54 & 0.86 & 0.04 \\
		{[O/Fe]} & 0.41 & 0.86 & 0.23 & 0.73 & 0.44 & 1.76 & 0.36 & 0.13 & 0.65 & 0.45 & 0.07 \\
		{[Na/Fe]} & 0.07 & -0.03 & 0.05 & 0.15 & 0.08 & 0.19 & 0.30 & 0.25 & -0.12 & 0.13 & 0.22 \\
		{[Al/Fe]} & 0.27 & 0.19 & 0.89 & 0.53 & 0.23 & 0.36 & 0.49 & 0.64 & 0.45 & 0.21 & 0.08 \\
		{[Mg/Fe]} & 0.38 & 0.29 & 0.15 & 0.48 & 0.62 & 0.20 & 0.27 & 0.20 & 0.63 & 0.28 & 0.05 \\
		{[Si/Fe]} & 0.14 & 0.19 & 0.17 & 0.52 & 0.28 & 0.17 & 0.21 & 0.29 & 0.40 & 0.08 & 0.02 \\
		{[Ca/Fe]} & -0.05 & 0.21 & 0.09 & 0.26 & 0.20 & 0.14 & 0.18 & 0.36 & 0.29 & 0.19 & 0.07 \\
		{[Ti/Fe]} & 0.10 & 0.01 & 0.13 & 0.35 & 0.35 & 0.18 & 0.25 & 0.30 & 0.45 & 0.20 & 0.04 \\
		{[Sc/Fe]} & 0.25 & -0.06 & 0.12 & 0.35 & 0.32 & 0.31 & 0.25 & 0.25 & 0.44 & 0.18 & 0.04 \\
		{[V/Fe]} & -0.08 & -0.14 & -0.15 & -0.01 & -0.08 & 0.08 & -0.07 & -0.12 & 0.13 & -0.18 & 0.06 \\
		{[Cr/Fe]} & 0.07 & -0.18 & -0.17 & -0.03 & -0.05 & - & -0.08 & -0.12 & 0.10 & 0.21 & 0.04 \\
		{[Mn/Fe]} & - & - & - & -0.35 & 0.13 & -0.40 & -0.41 & -0.52 & -0.06 & - & 0.08 \\
		{[Co/Fe]} & 0.08 & 0.18 & 0.16 & 0.14 & 0.13 & 0.22 & 0.15 & 0.14 & 0.24 & 0.05 & 0.05 \\
		{[Ni/Fe]} & -0.11 & 0.05 & -0.20 & -0.15 & 0.09 & -0.09 & -0.09 & -0.10 & 0.07 & -0.23 & 0.04 \\
		{[Cu/Fe]} & -1.00 & -0.35 & -0.76 & -0.95 & -0.68 & -0.46 & -0.57 & -0.76 & -0.60 & -0.85 & 0.05 \\
		{[Zn/Fe]} & - & - & - & -0.06 & -0.10 & -0.36 & -0.08 & -0.14 & -0.01 & - & 0.03 \\
		{[Zr/Fe]} & - & - & - & 0.13 & 0.28 & -0.22 & 0.14 & 0.23 & 0.61 & - & 0.04 \\
		{[Ba/Fe]} & 0.09 & -0.03 & -0.18 & 0.59 & 2.01 & -0.07 & 0.23 & 0.23 & 2.00 & -0.10 & 0.05 \\
		{[La/Fe]} & -0.11 & -0.32 & 0.19 & 0.38 & 1.38 & -0.05 & -1.20 & 0.06 & 0.96 & 0.17 & 0.05 \\
		{[Ce/Fe]} & - & - &  & 0.38 & 1.41 & 0.19 & 0.46 & 0.42 & 1.25 & - & 0.04 \\
		{[Y/Fe]} & - & - & -0.52 & 0.46 & 1.10 & -0.36 & 0.30 & 0.08 & 0.91 & -0.02 & 0.05 \\
		{[Eu/Fe]} & 0.20 & 0.29 & 0.53 & 0.32 & 0.76 & 0.48 & 0.35 & 0.34 & 0.77 & 0.25 & 0.06 \\
		{[Nd/Fe]} & 0.23 & 0.41 & 0.77 & 0.78 & 1.39 & 0.42 & 0.64 & 0.57 & 1.21 & 0.48 & 0.04 \\
		\noalign{\smallskip}\hline
	\end{tabular}
    \tablecomments{$^a$: Typical abundance error of each element.}
\end{table}

\begin{table}[htbp]
	\centering
	\caption{Observation information of 33 N-rich field stars studied in this work. }
	\label{tab:a2}
	\begin{tabular}{c c c c c c}
		\hline\hline\noalign{\smallskip}	
		Object Name & Gaia Source ID & Source & S/N & Observation Date \\
         &  &  & (around 6000 \AA) &  \\
		\noalign{\smallskip}\hline\noalign{\smallskip}
		$\mathrm{Num28}$ & 364641868732115584 & CFHT & 121 & 2020-11-30 \\
		$\mathrm{Num32}$ & 701587619382431488 & CFHT & 84 & 2020-12-04 \\
		$\mathrm{Num24}$ & 3165682645690499840 & CFHT & 82 & 2020-12-05 \\
		$\mathrm{Num7}$ & 646620280034177664 & CFHT & 77 & 2020-12-05 \\
		$\mathrm{Num63}$ & 1323227854226937984 & CFHT & 81 & 2021-08-28 \\
		$\mathrm{Num1}$ & 1366614235165728000 & CFHT & 96 & 2021-08-28 \\
		$\mathrm{Num37}$ & 3602560813461369984 & CFHT & 71 & 2023-06-27 \\
		$\mathrm{Num10}$ & 2100388163371964928 & CFHT & 61 & 2023-07-10 \\
		$\mathrm{Num9}$ & 3277155365059485696 & Magellan & 115 & 2020-02-26 \\
		$\mathrm{Num38}$ & 4395247566817046400 & Magellan & 171 & 2020-02-27 \\
		$\mathrm{Num47}$ & 3646613262223485184 & Magellan & 160 & 2020-02-28 \\
		$\mathrm{Num58}$ & 3628603811516372864 & Magellan & 93 & 2020-02-28 \\
		$\mathrm{Num62}$ & 3090695166699934080 & Magellan & 130 & 2020-02-27 \\
		$\mathrm{Num65}$ & 4417483269639370880 & Magellan & 152 & 2020-02-27 \\
		$\mathrm{Num67}$ & 2685646524818105344 & Magellan & 56 & 2019-07-24 \\
		$\mathrm{Num69}$ & 3835184248030417664 & Magellan & 149 & 2020-02-28 \\
		$\mathrm{Num80}$ & 576599363783579008 & Magellan & 129 & 2020-02-27 \\
		$\mathrm{Num82}$ & 3628877246314114176 & Magellan & 175 & 2020-02-28 \\
		$\mathrm{Num86}$ & 4455038291879056384 & Magellan & 51 & 2019-07-24 \\
		$\mathrm{Num88}$ & 4452741961844020992 & Magellan & 54 & 2019-07-24 \\
		$\mathrm{Num94}$ & 3866576748112550400 & Magellan & 62 & 2020-02-28 \\
		$\mathrm{Num97}$ & 1174906033445812352 & Magellan & 57 & 2019-07-24 \\
		$\mathrm{Num98}$ & 3733311094103785344 & Magellan & 184 & 2020-02-28 \\
        $\mathrm{G1}$ & 6759421390173629952 & GALAH & 80 & 2017-07-25 \\
        $\mathrm{G2}$ & 4421607950076317184 & GALAH & 88 & 2023-05-04 \\
        $\mathrm{G3}$ & 2613662391901002240 & GALAH & 87 & 2017-06-02 \\
        $\mathrm{G4}$ & 6623590781687449856 & GALAH & 108 & 2017-09-12 \\
        $\mathrm{G5}$ & 5494723292962635904 & GALAH & 138 & 2017-12-06 \\
        $\mathrm{G6}$ & 2447417402012872960 & GALAH & 172 & 2016-11-06 \\
        $\mathrm{G7}$ & 2618486563591431680 & GALAH & 126 & 2014-07-12 \\
        $\mathrm{G8}$ & 6049790293472210560 & GALAH & 184 & 2014-07-08 \\
        $\mathrm{G9}$ & 4664662314020413952 & GALAH & 118 & 2016-10-14 \\
        $\mathrm{G10}$ & 6049561049599634304 & GALAH & 77 & 2014-07-13 \\
		\noalign{\smallskip}\hline
	\end{tabular}
\end{table}

\begin{figure}[htbp]
	\centering
	\includegraphics[width=17cm,height=10.0cm]{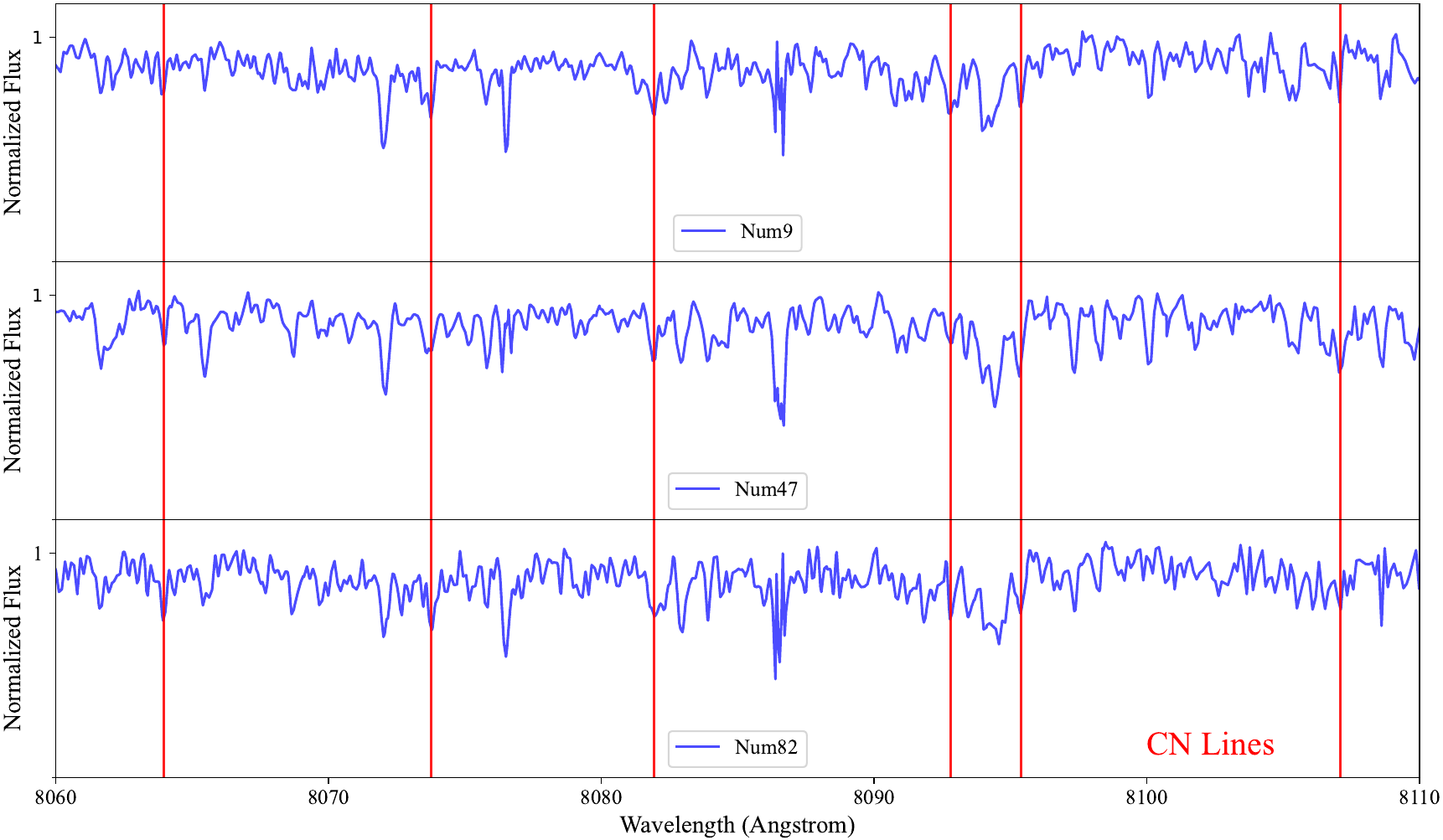}
	\caption{Sample spectra for 3 N-rich field stars studied in this work. Num47 and Num82 have converged C/N abundances and Num9 does not. The central wavelengths of feature CN lines are marked with red solid lines. }
	\label{figspec}
\end{figure}

\end{document}